\documentclass[twocolumn,prb,english]{revtex4-2}
\pdfoutput=1 
\usepackage{appendix}
\usepackage{dcolumn}
\usepackage{float}
\usepackage{amsmath,mathrsfs,bm,graphicx,xcolor}
\usepackage{hyperref}
\hypersetup{
     colorlinks = true,
     linkcolor  = blue,
     citecolor  = blue,  
     urlcolor   = blue, 
}
\begin{document}

\title{Transverse Peierls Transition}
  
\author{Kaifa Luo$^{1,2,4}$}
\author{Xi Dai$^{3,4}$}
\email{daix@ust.hk}
\affiliation{%
1, Department of Physics, University of Texas at Austin, Austin, Texas 78712, USA\\
2, Oden Institute for Computational Engineering and Sciences, University of Texas at Austin, Austin, Texas 78712, USA\\
3, Materials Department, University of California, Santa Barbara, California 93106, USA\\
4, Department of Physics, Hong Kong University of Science and Technology, Clear Water Bay, Hong Kong
}
\date{\today}

\begin{abstract}
In the present paper, we have discussed a new type of spontaneous symmetry breaking phases caused by the softening of the transverse acoustic phonon modes through the electron phonon coupling. These new phases include the shear density wave and self-twisting wave, which are caused by the softening of linearly and circularly polarized acoustic phonon modes, respectively. We propose that two of the topological semimetal systems in the quantum limit, where the electrons only occupy the lowest Landau bands under external magnetic field, will be the perfect systems to realise these new phases. Exotic physical effects will be induced in these new phases, including the 3D quantum Hall effect, chiral standing acoustic wave, magneto-acoustic effects and chiral phonon correction to Einstein-de Hass effect. 
\end{abstract}


\maketitle

\section{\label{sec:intro}Introduction}

The Peierls transition, which leads to charge density wave (CDW)\cite{RevModPhys.60.1129,Gruner_DW_1994} is one of the key phenomena caused by electron-phonon interaction\cite{grimvall1981electron,giustino2017electron} in condensed matter. 
It is induced by the so-called “nesting” feature of electron Fermi surfaces (FS), where two sections of the FS are connected by a single wave vector $Q$. 
The longitudinal acoustic (LA) phonon mode with wave vector $Q$ is strongly coupled to electron-hole excitation between different sections of the FS by electron-phonon interaction, which leads to singular response at the low temperature and causes ``condensation'' of that particular phonon mode. 
The Peierls transition to CDW usually happen in 1D materials like polyacetylene, where energy bands disperse along the chain direction and LA phonon with $Q=2k_{F}$ connecting two Fermi points condenses at low temperature. 
Another very different system shares similar Peierls' picture is 3D semimetal material\cite{RevModPhys.90.015001} under a strong magnetic field, where the electronic states are fully quantized to be Landau levels within the perpendicular plane and only disperse along the field direction. 
In the quantum limit, the Fermi level only crosses a single Landau band, which also satisfies the perfect nesting condition and leads to CDW. 
Comparing to 1D or quasi-1D materials, here the physics leading to completely flat band behaviour along the perpendicular directions is not the lack of overlap between the neighbouring electron wave functions but the Landau quantization. 
As a consequence, such a system has intrinsic 3D quantum Hall effect (QHE)\cite{Halperin1987,PhysRevB.45.13488} after CDW transition, which is quite different from the simple stacking of 2D quantum Hall layers\cite{stormer1986quantization}, because the Landau level spacing is overwhelmed by band dispersion in the former cases and 3D QHE only appears after a energy gap is open by interactions. 
Such a theoretical proposal of potential 3D QHE associated with CDW has been suggested for over three decades before it has finally been observed in Dirac material ZrTe$_5$\cite{tang2019three,qin2020theory}.

Comparing to QHE in 2D, much fruitful new physics will emerge in its 3D version. The most obvious difference is that 3D QHE is a spontaneous symmetry breaking phase under a strong magnetic field, which leads to new collective
dynamics originated from dynamics of the order parameters. 
For incommensurate CDW, which is the case in 3D QHE with generic field strength, such collective
modes are the ``phason'' modes along the field direction. 
How does the sliding motion of CDW induced by the phason modes couple with the quantum Hall physics will be an interesting problem to explore.

The CDW and Peierls transition discussed so far in the 1D metal and 3D Landau band systems are both caused by the condensation of LA phonon modes. 
What about the transverse acoustic (TA) phonon mode? Can TA phonon modes also couple to electrons and condense? 
What kind of new phases will be generated after condensation of TA phonon modes? 
These are key questions to be answered in the present paper. 
The results from our study reveal a completely different type of density wave instability associated with condensation of the TA phonon mode. 
Similar to the electromagnetic wave, TA wave is also a type of vector waves, which can carry angular momentum. 
Along some high-symmetry directions, conservation of total angular momentum $j_{z}\hbar$ will lead to ``selection rules'' in the electron-TA phonon interaction, which is the analog of optical selection rules. 
Interestingly, depending on the detailed features of the low energy electronic structure, the density waves caused by condensation of the TA phonons can be either linearly , elliptically or circularly polarized, which are also similar to the light. 
The corresponding linearly polarized density wave is a unique type of shear strain with periodical modulation, and the circularly polarized density wave can be viewed as self-twisting of the crystal along the external field direction.

As will be introduced in detail below, in the present work we have found two types of topological semimetals (TSMs) to realize the linearly  and circularly polarized TA phonon condensation under magnetic field, respectively. 
One class is Dirac semimetal with Dirac points being located along the high symmetry axis. 
Due to the existence of inversion symmetry, the stable density wave in this case is the ``shear strain wave'' caused by linearly polarized TA phonon condensation. 
The chiral TA phonon condensation may happen in another non-centrosymmetric system, Kramers-Weyl semimetal, which is found to be a perfect type of material to form ``self-twisting wave'' with the appearance of 3D QHE as its byproduct.

\section{Model}
Let us start with a generic model describing a semi-metal system with electron-phonon interaction,
    \begin{equation}\label{eq:model1}
    \begin{split}
    \hat{H} =
    &\frac{1}{N}\sum_{\alpha\beta\bm{k}}\mathcal{H}_{\bm{k}}^{\alpha\beta}\hat{c}_{\alpha\bm{k}}^{\dagger}\hat{c}_{\beta\bm{k}}
    +\frac{1}{N} \sum_{\lambda\bm{q}}\hbar\omega_{\lambda\bm{q}}\hat{b}_{\lambda\bm{q}}^{\dagger}\hat{b}_{\lambda\bm{q}}\\
    +&\frac{1}{N^{3/2}}\sum_{\alpha\beta\lambda\bm{k}\bm{q}}\mathcal{G}_{\bm{k}\bm{q}}^{\alpha\beta\lambda}
    \hat{c}_{\alpha\bm{k}+\bm{q}}^{\dagger}\hat{c}_{\beta\bm{k}}( \hat{b}_{\lambda\bm{q}} + \hat{b}_{\lambda, -\bm{q}}^{\dagger}) + h.c.,
    \end{split}
    \end{equation}
where $\mathcal{H}_{\bm{k}}$ is electron $k\cdot p$ Hamiltonian with $\alpha(\beta)$ being band index, $\omega_{\lambda\bm{q}}=v_{\lambda}^{ph}|\bm{q}|$ is the acoustic phonon frequency with polarization $\lambda=x,y,z$ and speeds $v_{\lambda}^{ph}$, and $ \mathcal{G}_{\bm{k}\bm{q}}$ is the electron-phonon interaction matrix. 
$\hat{c}_{\alpha\bm{k}}$ and $\hat{b}_{\lambda\bm{q}}$ are annihilation operators of electron and phonon, respectively.  
$N$ denotes the total number of unit cells. 
The system here has a discrete rotational symmetry $\hat{C}_{nz}$, and a strong external magnetic field $B_{z}\hat{\bm{z}}$ is applied. 
With respect to $\hat{C}_{nz}$, we adopt symmetric gauge $\bm{A} = (-y,x,0) B_{z}/2$ in cylindrical geometry and apply the Peierls substituion $k_{\pm}\rightarrow (-i\partial_{x}+eA_{x}/\hbar c)\pm i(-i\partial_{y}+eA_{y}/\hbar c)$ then. 
Under the magnetic field, in-plane motions of 3D electrons are fully quantized to form Landau levels and these ``Landau bands'' only disperse along $z$-direction. 
In the long wavelength limit, discrete rotation symmetry in crystals can be approximated to be continuous and the total angular momentum $j_{z}\hbar$ is conserved, which contains orbital part $l_{z}^{e}\hbar=(m-n)\hbar$ \cite{ezawa2008quantum} carried by the Landau level wave functions $|n,m\rangle$ ($n$ is Landau level index and $m$ the sub-index) and internal part $s_{z}\hbar$ (which is called ``spin'' here) inherited from the $k\cdot p$ model. 
Thus, all electronic states are uniquely labeled as $\hat{c}_{nmk_{z}}^{\dagger}$ hereafter. Note that, the effect of the direct coupling between charged ions with magnetic filed is weak enough to be ignored [Appendix \ref{app:chargedHO}].

Assuming that both the rotation and translation symmetries are along the $z$-axis, on linearly polarized basis, the canonical coordinates of the TA phonon modes can be expressed as $\hat{X}_{\pm q_{z}}=(\hat{X}_{x q_{z}}\pm i\hat{X}_{y q_{z}})/\sqrt{2}$, which carry the orbital angular momenta $\pm\hbar$. 
In the cases of TSMs with both electron-phonon interaction and strong SOC, spin $s_{z}$, electron orbital $l_{z}^{e}$ and phonon orbital $l_{z}^{ph}$ angular momenta are all coupled. However, $n$ index is fixed in the quantum limit, and $m$ is conserved. 
Thus, in our system, only $s_{z}$ and $l_{z}^{ph}$ is coupled [Appendix \ref{app:selectionRule}]. 
As a result, three selection rules are enforced to electron-phonon interaction due to translation $\hat{T}_{z}$ and rotation $\hat{C}_{z}$ symmetries:
    \begin{equation}\label{eq:SelectionRule}
    \begin{split}
    s_{z} = s_{z}' + l_{z,\lambda}^{ph},~m = m',~k_z = k_z' +q_z.
    \end{split}  
    \end{equation}

When $q,k\ll l_B^{-1}$, the forms of electron-phonon interaction for both LA and TA phonon modes can be derived in a unified way using the Bir-Pikus formalism \cite{Pikus_Bir_1974}, which describes the strain potential with both hydrostatic and shear deformation effects  included\cite{dumke1956deformation,Mahan_2010}. 
To be specific, a minimum non-trivial model with only two Landau bands is considered. 
Since these two bands of our interest will hold the same $n$ index, we drop it hereafter to lighten the notation. 
For a certain nesting wavevector $Q$, on a basis set 
$\hat{\Psi}_{mk_{z}Q}^{\dagger} = 
(\hat{c}_{s_{z}^{\alpha} mk_{z}+Q/2}^{\dagger}, 
\hat{c}_{s_{z}^{\beta} m k_{z}+Q/2}^{\dagger}, 
\hat{c}_{s_{z}^{\alpha}m k_{z}-Q/2}^{\dagger},
\hat{c}_{s_{z}^{\beta} m k_{z}-Q/2}^{\dagger})^T$, 
with only $\hat{C}_{z}$ imposed, the specific form of electron-phonon interaction up to the zeroth order of $k$ and first order of $q$ [Appendix \ref{app:B-2}] is,
    \begin{equation}\label{eq:epc}
    \begin{split}
    \mathcal{G}_{m k_{z}Q}^{\lambda=\pm} = 
        & iQ \xi_{\pm,Q}g_{\pm}\sigma_{\mp}\delta(s_{z}^{\alpha}-s_{z}^{\beta}-l_{z,\pm}^{ph}),\\
    \mathcal{G}_{m k_{z}Q}^{\lambda=z} = 
        & iQ \xi_{z Q}(g_{0}\sigma_0 + g_{z}\sigma_z)\delta(s_{z}^{\alpha}-s_{z}^{\beta}), 
    \end{split}
    \end{equation}
where Pauli matrices $\sigma_{\lambda}$ span the pseudo-spin space and $\sigma_{\pm}=(\sigma_{x}\pm i\sigma_{y})/2$, $\xi_{\lambda q_{z}}=\sqrt{\hbar/2M\omega_{\lambda q_{z}}}$ is the zero-point displacement amplitude with $M$ the mass of ions in each unit cell, and $g_{\lambda}$ is the coupling constant of the electron-phonon interaction. For simplicity, here we set $g_{0}=0$.

Due to the nesting feature of the FS, the ionic motion will couple very strongly to the electronic degree of freedoms. 
Such a system is unstable and can be dealt with by the mean-field approach to replace the canonical coordinate operators with their expectation values $\langle \hat{X}_{\lambda Q}\rangle$, 
then the mean-field Hamiltonian per unit-cell reads [Appendix \ref{app:C-3}],
    \begin{equation}
    \begin{split}\label{eq:mfHamil}
    &\hat{\bar{H}}_{Q}
        =\frac{1}{N}\sum_{mk_{z}}(\mathcal{H}_{mk_{z}Q}+\Delta_{mk_{z}Q})
            \hat{\Psi}^{\dagger}_{mk_{z}Q}\hat{\Psi}_{mk_{z}Q}\\
        &~~~~+\frac{M}{N}\sum_{\lambda}\left(g_{\lambda}^{-1}v_{\lambda}^{ph}|\Delta_{\lambda Q}|\right)^{2},\\
    \end{split}
    \end{equation}
where $\mathcal{H}_{m k_{z}Q}$ now becomes the effective Hamiltonian on the basis $\hat{\Psi}_{mk_{z}Q}^{\dagger}$ in the reduced Brillouin zone $k_{z}\in[-Q/2,Q/2]$. 
The ratio of shear to longitudinal wave speeds is determined by Poisson's ratio $\sigma$ through relation $v_{T}^{ph}/v_{z}^{ph}=\sqrt{(1-2\sigma)/(2-2\sigma)}$. 
The order parameter is a matrix, $\Delta_{mk_{z}Q}=\sum_{\lambda}\tau_{+}\mathcal{G}_{mk_{z}Q}^{\lambda}\Delta_{\lambda Q}/(iQ\xi_{\lambda Q}g_{\lambda})+h.c.$, and will be determined through self-consistent loop iteratively, where $\tau_{\pm}=(\tau_{x}\pm i\tau_{y})/2$ span the valley degree of freedom. 
In the following, we discuss two classes of realistic materials, Dirac and Kramers-Weyl semimetals.

    \begin{figure}[ht]
      \includegraphics[width=0.48\textwidth]{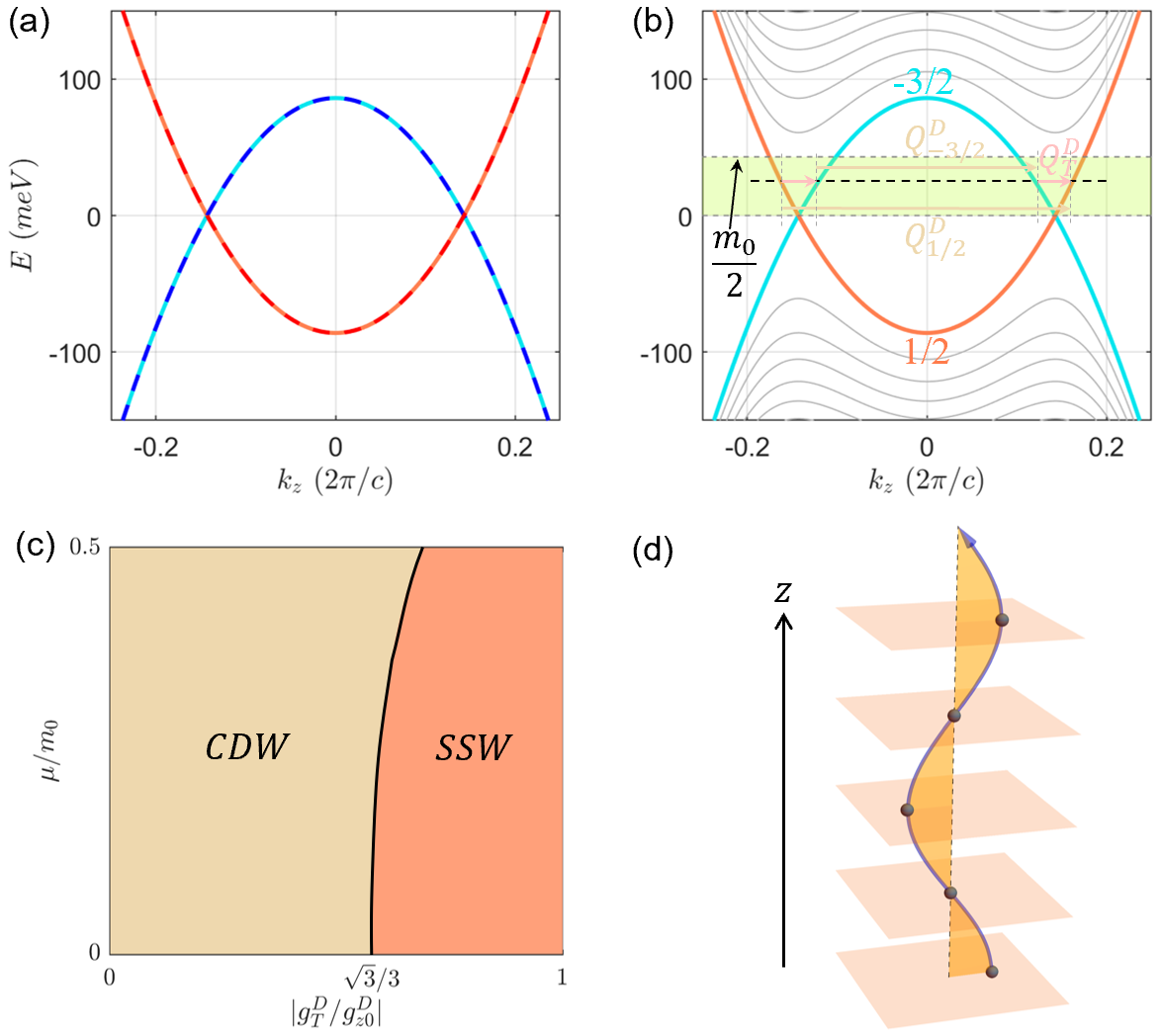}
      \caption{\label{Fig2}Dirac semimetal. 
      (a) Band structure near $\Gamma$ point in $k_z$-axis. 
      (b) Zeroth Landau bands when $B_{z}=10T$. The area in olive drab denotes the regime of tunable chemical potential $\mu$. 
      (c) The ground state is either CDW with LA phonon softened (regime in grey) or shear strain wave with linearly polarized TA phonon softened (regime in pink). When $\mu=0$, effective coupling strength $|g^{D}_{T}/g^{D}_{z}|=v_{T}^{ph}/v_{z}^{ph}=\sqrt{3}/3$ is the phase boundary while it shifts to the right when $\mu$ increases. 
      (d) Schematics of charge center (black dots) distribution in SSW phase.
      }
    \end{figure}

\section{Realistic Materials}

The first class of material considered in the present study is Dirac semimetal, such as Na$_{3}$Bi\cite{Wang2013-1,Wang2013-2,Liu2021,Liu2014}, in which a pair of Dirac points located at $k_{z}$-axis are protected by time reversal symmetry (TRS) $\hat{\mathcal{T}}$, inversion symmetry $\hat{\mathcal{P}}$ and rotation symmetry $\hat{\mathcal{C}}_{nz}$ ($n=3$ for Na$_{3}$Bi), as shown in Fig.\ref{Fig2}(a). 
Near $\Gamma$ point, the effective Hamiltonian reads 
$\mathcal{H}_{\bm{k}}^{D} = (m_z k_z^2 + m_{\perp} k_+ k_- - m_0) \tau_z \sigma_0 
+ \hbar v_{\perp} (\tau_x \sigma_z - \tau_y \sigma_0) - \mu$ 
on basis 
$\hat{\Psi}^{D\dagger}_{\bm{k}} = (\hat{c}_{1/2 \bm{k}}^{\dagger}, \hat{c}_{-1/2 \bm{k}}^{\dagger}, \hat{c}_{3/2 \bm{k}}^{\dagger}, \hat{c}_{-3/2 \bm{k}}^{\dagger})^T$ 
where electron states are denoted as $\hat{c}_{s_{z}\bm{k}}^{\dagger}$. 
After the magnetic field is applied, the low-energy physics is dominated by two zeroth Landau bands with spin $s_{z} = 1/2$ and $-3/2$ as shown in Fig.\ref{Fig2}(b). 
The basis becomes $\hat{c}_{s_{z} m k_{z}}$, and the effective magnetic Hamiltonian is 
$\mathcal{H}_{k_z}^{D} = (m_z k_z^2-m_0)\sigma_{z}-\mu$ [Appendix \ref{app:C-1}]. 
The FS contains four points (except $\mu=0$), corresponding to states $\hat{c}^{\dagger}_{s_{z},\pm k_{F,s_{z}}}$. 
Three phonons with different wave vectors are possible to participate in FS nesting, two inter-valley  $Q_{s_{z}}^{D}$ and one intra-valley $Q_{T}^{D}$.

\begin{table}[h]
    \caption{Model parameters for Dirac (Na$_{3}$Bi)\cite{Wang2013-2} and Kramers-Weyl ($\beta$-Ag$_{2}$Se)\cite{Wan2018} semimetals. The relative atomic mass $M_{0}=1.661\times 10^{-27}kg$ and Landé $g$-factor is set as $2$.}\label{tb:paras}
  \begin{tabular}{c|c|c|c|c}
    \hline\hline
    $m_0$     & $m_z$               & $m_{\perp}$         & $\hbar v_{\perp}$ &  $v_F^{D}$ \\
    0.087$eV$ & 10.64$eV\cdot\AA^2$ & 10.36$eV\cdot\AA^2$ & 2.46$eV\cdot\AA$  &  $289 km/s$ \\
    \hline
    $M^{D}/M_{0}$    & $g_{z0}^{D}$\cite{li2021charge} & $v_{z}^{ph,D}$ & $\sigma^{D}$ & \\
    556 & 0.5$eV$                        & $2650m/s$     & 0.25         &  \\
    \hline
    $u_z$            & $u_{\perp}$ & $v_z$           & $v_{\perp}$ & $v_{F}^{KW}$ \\
    -6$eV\cdot\AA^2$ & $2.5u_z$    & 0.2$eV\cdot\AA$ & 0.3$v_{z}$  & $v_z/\hbar$ \\
    \hline
    $M^{KW}/M_{0}$    & $g_{z0}^{KW}$\cite{shi2021charge} & $v_{z}^{ph,KW}$ & $\sigma^{KW}$ & \\
    1180 & 0.1$eV$                          & $2088m/s$      & 0.4           &  \\
    \hline\hline
  \end{tabular}
\end{table}

Enforced by the selection rules Eq.(\ref{eq:SelectionRule}), the LA phonon mode is only allowed to participate two  inter-valley scattering processes to form CDW phase. 
In contrast, the intra-valley scattering process involving electronic states with different $s_{z}$ can happen only when circularly polarized TA phonon modes ($l_{z}^{ph}\neq 0$) is considered. 
Depending on the electron-phonon interaction strength $g_{\lambda}$, the competition between LA and TA phonons results in different phases, which are described by the Hamiltonian Eq.(\ref{eq:mfHamil}). 
The condensation of the LA phonon mode leads to ordinary CDW order, which spontaneously breaks the translation symmetry only. 
The situation of TA phonon is quite different. 
Similar to electromagnetic wave, a TA phonon mode $\hat{X}_{\lambda q_{z}}$ has two polarizations, which can be expressed on circularly polarized basis $\hat{X}_{\pm,q_{z}}=(\hat{X}_{x,q_{z}}\pm i\hat{X}_{y,q_{z}})/\sqrt{2}$. 
When a TA phonon mode connects two different FS sections of the zeroth Landau bands as illustrated in Fig.\ref{Fig2}.(b), it can also be softened or even condensed. 
After condensation, in the generic case the expectation value of the phonon operators with both polarization will be nonzero. 
Interestingly, the condensation of the TA phonon modes can be either linear, elliptical and circular polarised depending on the relative phase factor $e^{i\phi_{xy}}=\langle \hat{X}_{y}\rangle/\langle \hat{X}_{x}\rangle$ between the order parameters of two orthogonal linearly polarized TA modes as summarised in the Table.(\ref{tb:phase}).

\begin{table}[ht]
    \caption{Three types of polarizations of ground states resulted by TA phonon condensation. Below $n$ is an integer.}\label{tb:phase}
  \begin{tabular}{c|c|c|c}
    \hline\hline
    Relative Phase & $\phi_{xy}=n\pi$ & $\phi_{xy}\neq n\pi/2$ & $\phi_{xy}=(2n+1)\pi/2$ \\
    Polarization & Linear & Elliptical & Circular \\
    \hline\hline
  \end{tabular}
\end{table}

In particular, the presence of inversion symmetry in Dirac semimetal will guarantee left- and right-handed TA modes to be condensed with the same amplitudes, corresponding to a linearly polarized TA mode actually. Unlike charge redistribution along the wave vector in LA phonon condensation, the condensation of linearly polarized TA phonon generates a shear strain with periodicity $Q_{T}^{D}$ instead, which is called as "shear strain wave" in our work. 
Based on the model parameters listed in Table.(\ref{tb:paras}), the mean-field phase diagram is plotted in Fig.\ref{Fig2}(c), in which CDW and shear strain wave phases are both possible to be stabilized in different regions of the parameter space. 
The key parameter determining the ground state is the ratio of coupling strength $|g_{T}^{D}/g_{z0}^{D}|$, which is the horizontal axis of the Fig.\ref{Fig2}(c). 
When it exceeds $v_{T}^{ph,D}/v_{z}^{ph,D}=\sqrt{3}/3$ at $\mu=0$, the shear strain wave phase is more stable than the CDW phase. 
Note that, after the CDW or self-twisting wave phases have been stabilized, the Dirac semimetal becoms a trivial insulator rather than 3D QHE state because the energy gap is opened on the zeroth Landau bands.

\begin{figure}[ht]
  \includegraphics[width=0.48\textwidth]{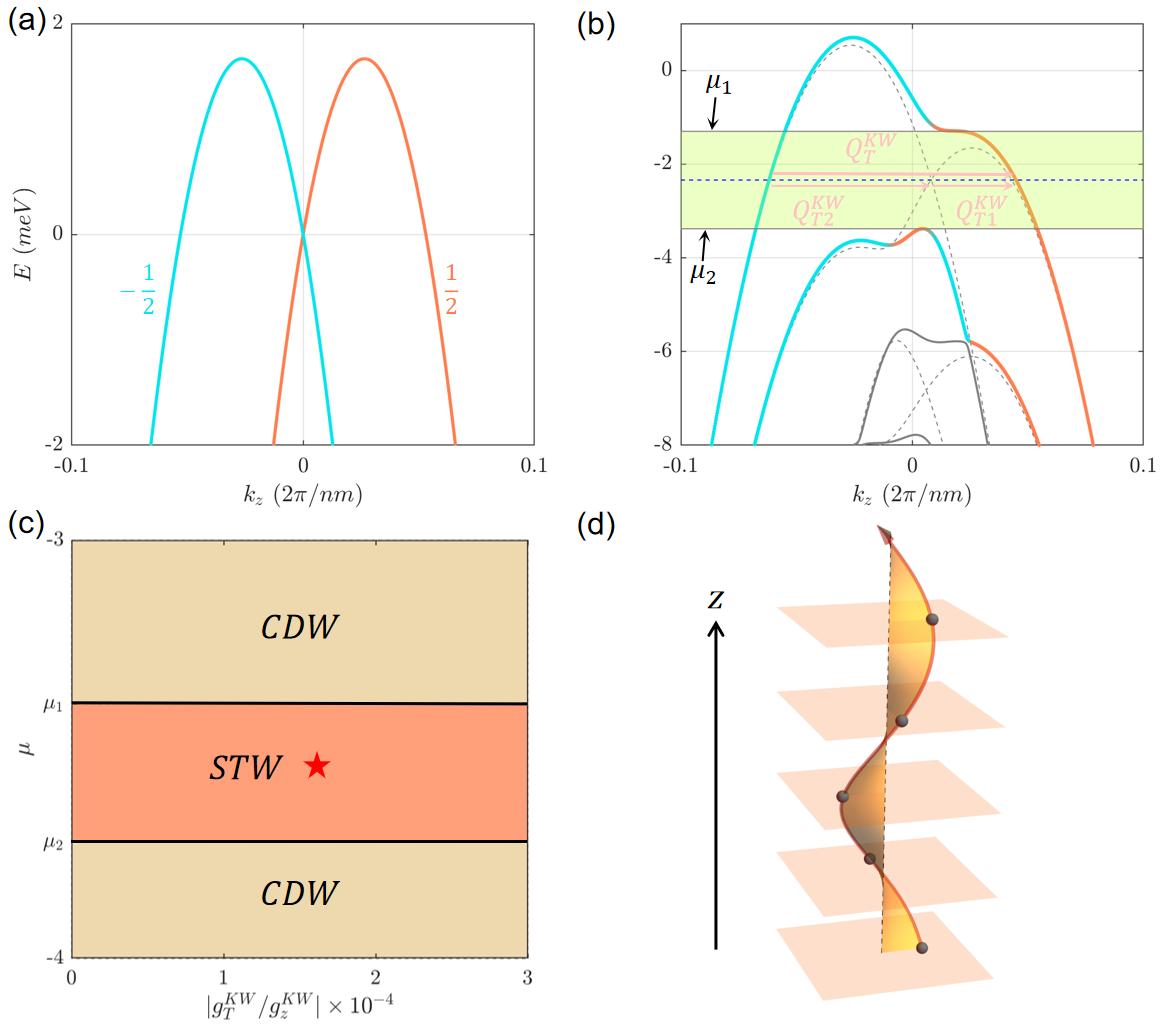}
  \caption{\label{Fig3} Kramers-Weyl semimetals. 
(a) Band structure near $\Gamma$ point in $k_z$ axis. 
(b) Lowest Landau bands when magnetic fields $B_{z}=10T$ and rotation symmetry breaking term $\delta_{T}$ are both applied, where the latter leads a gap $2\delta_{T}=\mu_{1}-\mu_{2}=2.4meV$ between $n=1$ and 2 Landau bands. 
The color shades in red and blue denoting proportions of wave functions with $s_{z} = \pm 1/2$, and they are mixed due to $C_{2z}$ breaking. 
(c) The phase diagram of ground state has two regimes, where the area in burnt orange: self-twisting wave (STW) when Fermi level lies within the gap (area in olive of (b)) while other regimes are CDW phase. The red star marks the parameters used to in Fig.(\ref{Fig4}). Here we set coupling constants $g_{+}=0.5g_{-}$. 
(d) Schematics of charge center (black dots) distribution in self-twisting phase.
}
\end{figure}

Another class is Kramers-Weyl semimetal such as $\beta$-Ag$_2$Se
\cite{Zhang2017,Chang2018,Wan2018}, which is not centrosymmetric and circularly polarized TA phonons will not be suppressed by $\hat{\mathcal{P}}$. 
Unlike ordinary Weyl semimetals, the Weyl points in Kramers-Weyl semimetal are all pinned at the time reversal invariant momenta. For example, the band structure around $\Gamma$ point is shown in Fig.\ref{Fig3}(a). 
Assuming that the system contains a $C_{nz}$($n=2$ in $\beta$-Ag$_2$Se) symmetry, its $k\cdot p$ model near $\Gamma$ point reads 
$\mathcal{H}_{\bm{k}}^{KW}= (u_z k_z + u_{\perp} \bm{k}_{\perp}^2) \sigma_0 +(v_z k_z \sigma_z + v_{\perp} \bm{k}_{\bm{\perp}} \cdot \bm{\sigma}_{\perp}) - \mu$  
on basis $\hat{\Psi}_{\bm{k}}^{KW \dagger}=(\hat{c}_{1/2\bm{k}}^{\dagger},\hat{c}_{-1/2\bm{k}}^{\dagger})^{T}$. 
Then in the quantum limit, the low-energy physics is dominated by two Landau bands with Landau band index $n=1$ and spin $s_{z}=\pm 1/2$, as shown in Fig.\ref{Fig3}(b)[Appendix \ref{app:C-2}].

To reach the nesting condition that the FS contains only two points, the Fermi level has to be placed inside the gap at the $\Gamma$ point, which is generated by an rotational symmetry breaking term in addition to the magnetic field. Such a rotation symmetry breaking term can be either generated by the Zeeman effect brought by an additional in-plane magnetic field or some kinds of strain. 
Strictly speaking, the wave functions of different Landau bands are mixed by the symmetry breaking term and angular momentum is no long a good quantum number. 
However, such mixing of states with different angular momenta is only significant near the $\Gamma$ point, while for states at the FS, the mixing effect is negligible and the angular momentum is still approximately conserved as well as the corresponding selection rule of electron-phonon interaction. 
Thus, in such a system the dominant instability happens for the circularly polarized TA phonons because of the selection rule. 
As a result, electron-phonon interaction will lead to self-twisting wave phase followed by the condensation of chiral TA phonon modes, as shown in Fig.\ref{Fig4}(c). 
Since now the gap opens for the Landau bands with index $n=1$, the electronic ground state will have 3D QHE with quantized transverse conductance $\sigma_{STW} = Q_{T}^{KW}e^{2}/h$.
                                                                                                                                                                   
\section{Physical Effects}

In this section, we discuss the possible exotic physical phenomena caused by the condensation or softening of the TA phonon modes. 
The first is the Goldstone mode in self-twisting wave phase. Similar to CDW phase where the sliding mode is a super current of charge as a result of spontaneous breaking of translation symmetry, the chiral sliding mode in the incommensurate self-twisting wave phase is a super current of angular momentum instead. Also, the new ground state will be a perfect chiral crystal, where the chiral phonon is discussed in a recent work\cite{chen2021chiral}.

\begin{figure}[ht]
  \includegraphics[width=0.48\textwidth]{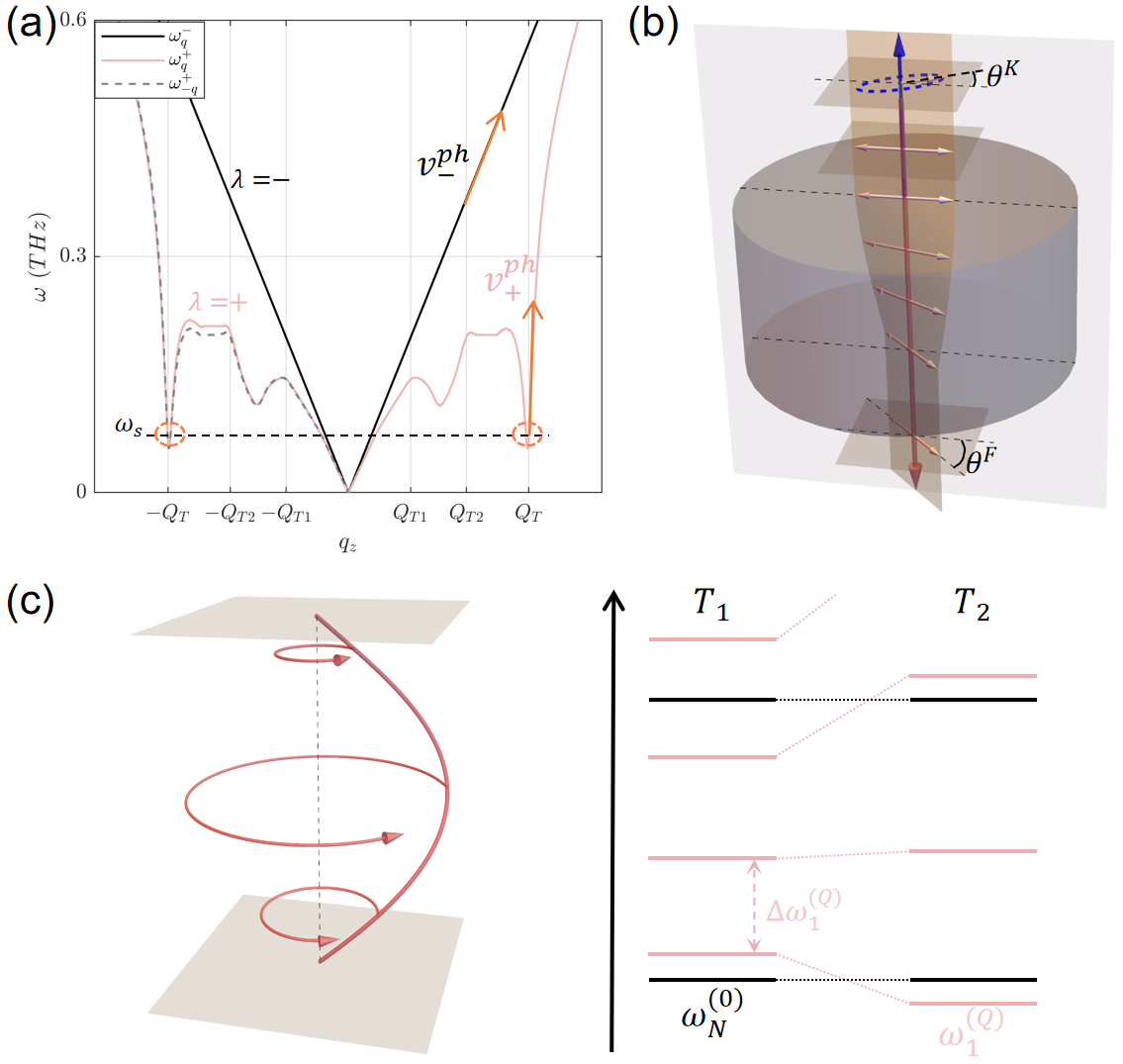}
  \caption{\label{Fig4} Kramers-Weyl semimetals. (a) Renormalized TA phonon frequencies at $T=1.1K$ (near $T_{c}\approx 1K$). Here $|g_{T}/g_{z0}|=1.43\times 10^{-4}$ and the speeds of left-handed (pink) and right-handed (black) chiral TA phonons are $v_{\pm}^{ph}$. $\omega_{s}$ is the frequency of standing acoustic waves.
  (b) Magneto-acoustic effects. The gyromagnetic acoutic medium (cylinder in grey) is surrounded by regular medium. When a linearly polarized TA wave (the two-head arrow) incident to the interface normally, the reflected wave (blue) becomes elliptical with Kerr elliplicity $\epsilon^{K}$ and rotation angle $\theta^{K}$, while the polarization plane of refracted wave (red) gradually rotates by Faraday angle $\theta^{F}$ when it moves forward. 
  (c) Schematics of left-handed chiral standing acoustic wave between two nodes (left), and resonant frequencies of standing modes (right). Given a finite length system at $T_{1}=1.1K$, the frequencies of chiral standing modes with wave vector near $Q_{T}$ is $\omega_{Q_{T}}+\Delta\omega_{n}^{(Q_{T})}$, and of normal long wavelength standing modes (black) is $\omega_{n'}^{(0)}$. $\omega_{N}^{(0)}$ is the $N$-th normal standing mode close to $\omega_{1}^{(Q_{T})}$. When the system is cooled down from $T_{1}$ to $T_{2}$ ($\sim 0.02K$), the frequencies of $\omega_{n}^{(0)}$ are almost the same, while the fundamental chiral mode $\omega_{Q_{T}}$ decreases rapidly and the frequency differences $\Delta\omega_{n}^{(Q_{T})}$ increase.
  }
\end{figure}

More interesting effects are from acoustical activity\cite{PhysRev.170.673,frenzel2019ultrasound,luthi2007physical}. To analyze frequency renormalization, we consider the leading order of phonon self-energy [Appendix \ref{app:E}],
\begin{equation}\begin{aligned}
  \Delta\omega^{2}_{\lambda q_{z}}
  =&(\omega_{\lambda q_{z}}^{ren})^{2} 
  - \omega_{\lambda q_{z}}^{2}=\frac{(g_{T}q_{z})^2}{M}\mathscr{L}_{\lambda q_z}(\omega_{\lambda q_{z}}^{ren}),\\
  \mathscr{L}_{\lambda q_{z}}
  =&\frac{1}{N}\sum_{\alpha\beta m k_{z}}
  \frac{(f_{\alpha k_z+q_z}-f_{\beta k_z})\delta(s_{z}^{\alpha}-s_{z}^{\beta}-l_{z,\lambda}^{ph})}{\varepsilon_{\alpha k_z+q_z} - \varepsilon_{\beta k_z}-\hbar(\omega+i\eta)},
\end{aligned}\end{equation}
where $\mathscr{L}_{\lambda q_{z}}$ is the Lindhard response function with $f_{\alpha k_{z}}\equiv f(\varepsilon_{\alpha k_{z}})$ being Fermi-Dirac distribution. 
At low temperature $T<\delta_{T}/k_{B}$ where the gap will not be smeared out by thermal fluctuation, the TA phonon frequencies at long wavelength regime is almost not affected and quasi-degenerate, while a huge deviation between two different chiral phonons happens near $Q_{T}$ wave vector regime and effective dispersion of the left-handed phonon branch near $\pm Q_{T}$ is $\omega_{\delta q_{z}}^{(\pm Q_{T})}=\omega_{Q_{T}}\pm a\delta q_{z}+b(\delta q_{z})^{2}/2$ with $\omega_{Q_{T}}=v_{T}^{ph}Q_{T}\sqrt{(T-T_{c})/T_{c}}$ before phase transition happens ($T$ higher than critical temperature $T_{c}$), and coefficients $a=1350m/s,b=9.1\times 10^{-4}m^{2}/s$ are fitted numerically at a certain temperature $T=1.1K$. 
Thus, the system is a gyromagnetic medium with acoustical activity, in which the group speed of right-handed phonon branch is still $v_{T}^{ph}$, and of left-handed chiral waves is renormalized to $v_{\delta q_{z}}^{(\pm Q_{T})}=\partial\omega_{\delta q_{z}}^{(\pm Q_{T})}/\partial(\delta q_{z})=\pm a+b\delta q_{z}$, as shown in Fig.\ref{Fig4}(a). 

 After the degeneracy of TA phonon modes is lifted by electron-phonon interaction, a series of magneto-acoustic effects are immediately followed, such as Faraday, Kerr, Cotton–Mouton or Voigt effects\cite{zvezdin1997modern,luthi2007physical}. As examples, here we discuss Faraday rotation and Kerr ellipticity as results of magneto-acoustic circular birefringence, which are also found in superfluid $^3$He–B\cite{lee1999discovery,sauls2000magneto}. 
 Consider a setup, in which the gyromagnetic acoustic medium is surrounded by normal medium such as chlorinated polyvinyl chloride (CPVC) with shear acoustic wave speed 1060m/s. 
 A linearly polarized TA wave with frequency $\omega\ge\omega_{Q}$ incident from CPVC side towards the gyromagnetic medium in polar configuration, the reflected wave from the interface will become elliptical, while the polarization plane of refracted wave in gyromagnetic medium will gradually rotate, as shown in Fig.\ref{Fig4}(b). 
Under external magnetic induction $B_{z}=10T$ at $T=1.1K$, for acoustic wave with frequency $\omega=58.0GHz$, a significant Kerr ellipticity $\epsilon^{K}=0.41$, and a huge Faraday angle per distance $1.77\times 10^{5} rad/cm$ [Appendix \ref{app:G}] are estimated to be measured. Other typical effects, such as Faraday ellipticity and Kerr rotation induced by magneto-acoustic circular dichroism can also be observed. 

Another easily observed signature by magneto-acoustic circular refringence is resonant frequency of chiral standing acoustic wave, as shown in the Fig.\ref{Fig4}(c). Given a system at $T_{1}=1.1K$ with length $d=1cm$ along $z$-axis, the resonant frequencies of the left-handed chiral modes with wave vector around $Q_{T}$ are $\omega_{Q_{T}}+\Delta\omega_{n}^{(Q_{T})}=56.1GHz+0.086(n+1/2)kHz$ and of normal long wavelength standing modes are $\omega_{n}^{(0)}=(268n)kHz$ [Appendix \ref{app:F}]. Because the electronic screening leads to the frequency local minimum $\omega_{Q_{T}}\propto\sqrt{(T-T_{c})/T_{c}}$\cite{Gruner_DW_1994}, a tiny temperature change will lead to a significant lowering of fundamental model of the chiral standing modes as well as increase of frequency differences. 
For example, $\omega_{Q}$ will decrease to $48.1GHz$ and $\Delta\omega_{n}^{(Q_{T})}$ increase to $0.097(n+1/2)kHz$ if the system is cooled down to $T=1.08K$. 
Similar temperature sensitive resonances can also be observed in shear strain wave phase of Dirac semimetal, while the standing modes are not chiral but linearly polarized.

In the end, we would also mention the chiral phonon correction to the Einstein-de Hass effect. 
When electron-phonon interaction is turned off and the direct coupling of ions to magnetic field is neglected, TRS will pair two phonons with opposite angular momenta as $\hat{\mathcal{T}}\hat{X}_{+,\bm{q}}\hat{\mathcal{T}}^{-1}=\hat{X}_{-,-\bm{q}}$, resulting in a zero net phonon angular momenta, which is one of the premises in original proposal of the Einstein-de Hass effect\cite{Einstein1915}. 
However, the TRS of TA phonon subsystem is broken mediated by the electron-phonon interaction. As the same result of spin-phonon interaction proposed by Zhang and Niu\cite{Zhang2014}, the non-zero net phonon angular momentum will lead to an observable correction to the Einstein-de Hass effect. 

\section{Conclusion and Outlook}

The interaction between electron and TA phonon mode has been overlooked for decades. 
In this article, we have proposed two kinds of exotic condensation of TA phonon modes, shear strain wave and self-density wave, induced by electron-TA phonon interaction in specific TSMs under magnetic field. We find that when the strong enough magnetic field is applied along the rotational axis of Dirac and Kramers-Weyl semimetal materials, the low energy electronic states can be described by their zeroth or lowest Landau bands with different internal angular momentum $s_z\hbar$. The presence of rotational symmetry enforces important selection rules to electron-phonon interaction terms for both LA and TA phonons. The combination of FS nesting and the selection rules of electron-phonon interaction induce two exotic new density wave states. The linearly polarized TA phonon condensation leads to shear strain wave state while circularly polarized TA phonon condensation gives us self-twisting state. The self-twisting phase is a typical chiral matter, which can lead to a number of new physical effects including chiral sliding mode, chiral standing wave, magneto-acoustic effects and chiral phonon correction to Eistein-de Has effect.


\subsection*{Acknowledgement}
We thank Qian Niu and Yafei Ren for valuable discussions. X. D. acknowledges financial support from the Hong Kong Research Grants Council (Project No. GRF16300918 and No. 16309020). K. L. acknowledges HPC resources from the Texas Advanced Computing Center (TACC) at The University
of Texas at Austin\cite{TACC}.
\bibliography{maintext.bib}

\begin{thebibliography}{33}%
\makeatletter
\providecommand \@ifxundefined [1]{%
 \@ifx{#1\undefined}
}%
\providecommand \@ifnum [1]{%
 \ifnum #1\expandafter \@firstoftwo
 \else \expandafter \@secondoftwo
 \fi
}%
\providecommand \@ifx [1]{%
 \ifx #1\expandafter \@firstoftwo
 \else \expandafter \@secondoftwo
 \fi
}%
\providecommand \natexlab [1]{#1}%
\providecommand \enquote  [1]{``#1''}%
\providecommand \bibnamefont  [1]{#1}%
\providecommand \bibfnamefont [1]{#1}%
\providecommand \citenamefont [1]{#1}%
\providecommand \href@noop [0]{\@secondoftwo}%
\providecommand \href [0]{\begingroup \@sanitize@url \@href}%
\providecommand \@href[1]{\@@startlink{#1}\@@href}%
\providecommand \@@href[1]{\endgroup#1\@@endlink}%
\providecommand \@sanitize@url [0]{\catcode `\\12\catcode `\$12\catcode
  `\&12\catcode `\#12\catcode `\^12\catcode `\_12\catcode `\%12\relax}%
\providecommand \@@startlink[1]{}%
\providecommand \@@endlink[0]{}%
\providecommand \url  [0]{\begingroup\@sanitize@url \@url }%
\providecommand \@url [1]{\endgroup\@href {#1}{\urlprefix }}%
\providecommand \urlprefix  [0]{URL }%
\providecommand \Eprint [0]{\href }%
\providecommand \doibase [0]{https://doi.org/}%
\providecommand \selectlanguage [0]{\@gobble}%
\providecommand \bibinfo  [0]{\@secondoftwo}%
\providecommand \bibfield  [0]{\@secondoftwo}%
\providecommand \translation [1]{[#1]}%
\providecommand \BibitemOpen [0]{}%
\providecommand \bibitemStop [0]{}%
\providecommand \bibitemNoStop [0]{.\EOS\space}%
\providecommand \EOS [0]{\spacefactor3000\relax}%
\providecommand \BibitemShut  [1]{\csname bibitem#1\endcsname}%
\let\auto@bib@innerbib\@empty
\bibitem [{\citenamefont {Gr\"uner}(1988)}]{RevModPhys.60.1129}%
  \BibitemOpen
  \bibfield  {author} {\bibinfo {author} {\bibfnamefont {G.}~\bibnamefont
  {Gr\"uner}},\ }\bibfield  {title} {\bibinfo {title} {The dynamics of
  charge-density waves},\ }\href {https://doi.org/10.1103/RevModPhys.60.1129}
  {\bibfield  {journal} {\bibinfo  {journal} {Rev. Mod. Phys.}\ }\textbf
  {\bibinfo {volume} {60}},\ \bibinfo {pages} {1129} (\bibinfo {year}
  {1988})}\BibitemShut {NoStop}%
\bibitem [{\citenamefont {Grüner}(1994)}]{Gruner_DW_1994}%
  \BibitemOpen
  \bibfield  {author} {\bibinfo {author} {\bibnamefont {Grüner}},\ }\bibfield
  {title} {\bibinfo {title} {Density waves in solids},\ }\href@noop {}
  {\bibfield  {journal} {\bibinfo  {journal} {CRC Press}\ } (\bibinfo {year}
  {1994})}\BibitemShut {NoStop}%
\bibitem [{\citenamefont {Grimvall}\ \emph {et~al.}(1981)\citenamefont
  {Grimvall} \emph {et~al.}}]{grimvall1981electron}%
  \BibitemOpen
  \bibfield  {author} {\bibinfo {author} {\bibfnamefont {G.}~\bibnamefont
  {Grimvall}} \emph {et~al.},\ }\href@noop {} {\emph {\bibinfo {title} {The
  electron-phonon interaction in metals}}}\ (\bibinfo {year}
  {1981})\BibitemShut {NoStop}%
\bibitem [{\citenamefont {Giustino}(2017)}]{giustino2017electron}%
  \BibitemOpen
  \bibfield  {author} {\bibinfo {author} {\bibfnamefont {F.}~\bibnamefont
  {Giustino}},\ }\bibfield  {title} {\bibinfo {title} {Electron-phonon
  interactions from first principles},\ }\href@noop {} {\bibfield  {journal}
  {\bibinfo  {journal} {Reviews of Modern Physics}\ }\textbf {\bibinfo {volume}
  {89}},\ \bibinfo {pages} {015003} (\bibinfo {year} {2017})}\BibitemShut
  {NoStop}%
\bibitem [{\citenamefont {Armitage}\ \emph {et~al.}(2018)\citenamefont
  {Armitage}, \citenamefont {Mele},\ and\ \citenamefont
  {Vishwanath}}]{RevModPhys.90.015001}%
  \BibitemOpen
  \bibfield  {author} {\bibinfo {author} {\bibfnamefont {N.~P.}\ \bibnamefont
  {Armitage}}, \bibinfo {author} {\bibfnamefont {E.~J.}\ \bibnamefont {Mele}},\
  and\ \bibinfo {author} {\bibfnamefont {A.}~\bibnamefont {Vishwanath}},\
  }\bibfield  {title} {\bibinfo {title} {Weyl and dirac semimetals in
  three-dimensional solids},\ }\href
  {https://doi.org/10.1103/RevModPhys.90.015001} {\bibfield  {journal}
  {\bibinfo  {journal} {Rev. Mod. Phys.}\ }\textbf {\bibinfo {volume} {90}},\
  \bibinfo {pages} {015001} (\bibinfo {year} {2018})}\BibitemShut {NoStop}%
\bibitem [{\citenamefont {Halperin}(1987)}]{Halperin1987}%
  \BibitemOpen
  \bibfield  {author} {\bibinfo {author} {\bibfnamefont {B.~I.}\ \bibnamefont
  {Halperin}},\ }\bibfield  {title} {\bibinfo {title} {Possible states for a
  three-dimensional electron gas in a strong magnetic field},\ }\href@noop {}
  {\bibfield  {journal} {\bibinfo  {journal} {Japanese Journal of Applied
  Physics}\ }\textbf {\bibinfo {volume} {26}},\ \bibinfo {pages} {1913}
  (\bibinfo {year} {1987})}\BibitemShut {NoStop}%
\bibitem [{\citenamefont {Kohmoto}\ \emph {et~al.}(1992)\citenamefont
  {Kohmoto}, \citenamefont {Halperin},\ and\ \citenamefont
  {Wu}}]{PhysRevB.45.13488}%
  \BibitemOpen
  \bibfield  {author} {\bibinfo {author} {\bibfnamefont {M.}~\bibnamefont
  {Kohmoto}}, \bibinfo {author} {\bibfnamefont {B.~I.}\ \bibnamefont
  {Halperin}},\ and\ \bibinfo {author} {\bibfnamefont {Y.-S.}\ \bibnamefont
  {Wu}},\ }\bibfield  {title} {\bibinfo {title} {Diophantine equation for the
  three-dimensional quantum hall effect},\ }\href
  {https://doi.org/10.1103/PhysRevB.45.13488} {\bibfield  {journal} {\bibinfo
  {journal} {Phys. Rev. B}\ }\textbf {\bibinfo {volume} {45}},\ \bibinfo
  {pages} {13488} (\bibinfo {year} {1992})}\BibitemShut {NoStop}%
\bibitem [{\citenamefont {St{\"o}rmer}\ \emph {et~al.}(1986)\citenamefont
  {St{\"o}rmer}, \citenamefont {Eisenstein}, \citenamefont {Gossard},
  \citenamefont {Wiegmann},\ and\ \citenamefont
  {Baldwin}}]{stormer1986quantization}%
  \BibitemOpen
  \bibfield  {author} {\bibinfo {author} {\bibfnamefont {H.}~\bibnamefont
  {St{\"o}rmer}}, \bibinfo {author} {\bibfnamefont {J.}~\bibnamefont
  {Eisenstein}}, \bibinfo {author} {\bibfnamefont {A.}~\bibnamefont {Gossard}},
  \bibinfo {author} {\bibfnamefont {W.}~\bibnamefont {Wiegmann}},\ and\
  \bibinfo {author} {\bibfnamefont {K.}~\bibnamefont {Baldwin}},\ }\bibfield
  {title} {\bibinfo {title} {Quantization of the hall effect in an anisotropic
  three-dimensional electronic system},\ }\href@noop {} {\bibfield  {journal}
  {\bibinfo  {journal} {Physical review letters}\ }\textbf {\bibinfo {volume}
  {56}},\ \bibinfo {pages} {85} (\bibinfo {year} {1986})}\BibitemShut {NoStop}%
\bibitem [{\citenamefont {Tang}\ \emph {et~al.}(2019)\citenamefont {Tang},
  \citenamefont {Ren}, \citenamefont {Wang}, \citenamefont {Zhong},
  \citenamefont {Schneeloch}, \citenamefont {Yang}, \citenamefont {Yang},
  \citenamefont {Lee}, \citenamefont {Gu}, \citenamefont {Qiao} \emph
  {et~al.}}]{tang2019three}%
  \BibitemOpen
  \bibfield  {author} {\bibinfo {author} {\bibfnamefont {F.}~\bibnamefont
  {Tang}}, \bibinfo {author} {\bibfnamefont {Y.}~\bibnamefont {Ren}}, \bibinfo
  {author} {\bibfnamefont {P.}~\bibnamefont {Wang}}, \bibinfo {author}
  {\bibfnamefont {R.}~\bibnamefont {Zhong}}, \bibinfo {author} {\bibfnamefont
  {J.}~\bibnamefont {Schneeloch}}, \bibinfo {author} {\bibfnamefont {S.~A.}\
  \bibnamefont {Yang}}, \bibinfo {author} {\bibfnamefont {K.}~\bibnamefont
  {Yang}}, \bibinfo {author} {\bibfnamefont {P.~A.}\ \bibnamefont {Lee}},
  \bibinfo {author} {\bibfnamefont {G.}~\bibnamefont {Gu}}, \bibinfo {author}
  {\bibfnamefont {Z.}~\bibnamefont {Qiao}}, \emph {et~al.},\ }\bibfield
  {title} {\bibinfo {title} {Three-dimensional quantum hall effect and
  metal--insulator transition in zrte 5},\ }\href@noop {} {\bibfield  {journal}
  {\bibinfo  {journal} {Nature}\ }\textbf {\bibinfo {volume} {569}},\ \bibinfo
  {pages} {537} (\bibinfo {year} {2019})}\BibitemShut {NoStop}%
\bibitem [{\citenamefont {Qin}\ \emph {et~al.}(2020)\citenamefont {Qin},
  \citenamefont {Li}, \citenamefont {Du}, \citenamefont {Wang}, \citenamefont
  {Zhang}, \citenamefont {Yu}, \citenamefont {Lu}, \citenamefont {Xie} \emph
  {et~al.}}]{qin2020theory}%
  \BibitemOpen
  \bibfield  {author} {\bibinfo {author} {\bibfnamefont {F.}~\bibnamefont
  {Qin}}, \bibinfo {author} {\bibfnamefont {S.}~\bibnamefont {Li}}, \bibinfo
  {author} {\bibfnamefont {Z.}~\bibnamefont {Du}}, \bibinfo {author}
  {\bibfnamefont {C.}~\bibnamefont {Wang}}, \bibinfo {author} {\bibfnamefont
  {W.}~\bibnamefont {Zhang}}, \bibinfo {author} {\bibfnamefont
  {D.}~\bibnamefont {Yu}}, \bibinfo {author} {\bibfnamefont {H.-Z.}\
  \bibnamefont {Lu}}, \bibinfo {author} {\bibfnamefont {X.}~\bibnamefont
  {Xie}}, \emph {et~al.},\ }\bibfield  {title} {\bibinfo {title} {Theory for
  the charge-density-wave mechanism of 3d quantum hall effect},\ }\href@noop {}
  {\bibfield  {journal} {\bibinfo  {journal} {Physical Review Letters}\
  }\textbf {\bibinfo {volume} {125}},\ \bibinfo {pages} {206601} (\bibinfo
  {year} {2020})}\BibitemShut {NoStop}%
\bibitem [{\citenamefont {Ezawa}(2008)}]{ezawa2008quantum}%
  \BibitemOpen
  \bibfield  {author} {\bibinfo {author} {\bibfnamefont {Z.~F.}\ \bibnamefont
  {Ezawa}},\ }\href@noop {} {\emph {\bibinfo {title} {Quantum Hall effects:
  Field theoretical approach and related topics}}}\ (\bibinfo  {publisher}
  {World Scientific Publishing Company},\ \bibinfo {year} {2008})\BibitemShut
  {NoStop}%
\bibitem [{\citenamefont {Bir}\ and\ \citenamefont
  {Pikus}(1974)}]{Pikus_Bir_1974}%
  \BibitemOpen
  \bibfield  {author} {\bibinfo {author} {\bibfnamefont {G.~L.}\ \bibnamefont
  {Bir}}\ and\ \bibinfo {author} {\bibfnamefont {G.~E.}\ \bibnamefont
  {Pikus}},\ }\href@noop {} {\emph {\bibinfo {title} {Symmetry and
  strain-induced effects in semiconductors}}},\ Vol.\ \bibinfo {volume} {624}\
  (\bibinfo  {publisher} {Wiley New York},\ \bibinfo {year} {1974})\BibitemShut
  {NoStop}%
\bibitem [{\citenamefont {Dumke}(1956)}]{dumke1956deformation}%
  \BibitemOpen
  \bibfield  {author} {\bibinfo {author} {\bibfnamefont {W.~P.}\ \bibnamefont
  {Dumke}},\ }\bibfield  {title} {\bibinfo {title} {Deformation potential
  theory for n-type ge},\ }\href@noop {} {\bibfield  {journal} {\bibinfo
  {journal} {Physical Review}\ }\textbf {\bibinfo {volume} {101}},\ \bibinfo
  {pages} {531} (\bibinfo {year} {1956})}\BibitemShut {NoStop}%
\bibitem [{\citenamefont {Mahan}(2013)}]{Mahan_2010}%
  \BibitemOpen
  \bibfield  {author} {\bibinfo {author} {\bibfnamefont {G.~D.}\ \bibnamefont
  {Mahan}},\ }\href@noop {} {\emph {\bibinfo {title} {Many-particle physics}}}\
  (\bibinfo  {publisher} {Springer Science \& Business Media},\ \bibinfo {year}
  {2013})\BibitemShut {NoStop}%
\bibitem [{\citenamefont {Wang}\ \emph {et~al.}(2013)\citenamefont {Wang},
  \citenamefont {Weng}, \citenamefont {Wu}, \citenamefont {Dai},\ and\
  \citenamefont {Fang}}]{Wang2013-1}%
  \BibitemOpen
  \bibfield  {author} {\bibinfo {author} {\bibfnamefont {Z.}~\bibnamefont
  {Wang}}, \bibinfo {author} {\bibfnamefont {H.}~\bibnamefont {Weng}}, \bibinfo
  {author} {\bibfnamefont {Q.}~\bibnamefont {Wu}}, \bibinfo {author}
  {\bibfnamefont {X.}~\bibnamefont {Dai}},\ and\ \bibinfo {author}
  {\bibfnamefont {Z.}~\bibnamefont {Fang}},\ }\bibfield  {title} {\bibinfo
  {title} {Three-dimensional dirac semimetal and quantum transport in cd 3 as
  2},\ }\href {https://doi.org/10.1103/PhysRevB.88.125427} {\bibfield
  {journal} {\bibinfo  {journal} {PHYSICAL REVIEW B}\ }\textbf {\bibinfo
  {volume} {88}},\ \bibinfo {pages} {125427} (\bibinfo {year}
  {2013})}\BibitemShut {NoStop}%
\bibitem [{\citenamefont {Wang}\ \emph {et~al.}(2012)\citenamefont {Wang},
  \citenamefont {Sun}, \citenamefont {Chen}, \citenamefont {Franchini},
  \citenamefont {Xu}, \citenamefont {Weng}, \citenamefont {Dai},\ and\
  \citenamefont {Fang}}]{Wang2013-2}%
  \BibitemOpen
  \bibfield  {author} {\bibinfo {author} {\bibfnamefont {Z.}~\bibnamefont
  {Wang}}, \bibinfo {author} {\bibfnamefont {Y.}~\bibnamefont {Sun}}, \bibinfo
  {author} {\bibfnamefont {X.-Q.}\ \bibnamefont {Chen}}, \bibinfo {author}
  {\bibfnamefont {C.}~\bibnamefont {Franchini}}, \bibinfo {author}
  {\bibfnamefont {G.}~\bibnamefont {Xu}}, \bibinfo {author} {\bibfnamefont
  {H.}~\bibnamefont {Weng}}, \bibinfo {author} {\bibfnamefont {X.}~\bibnamefont
  {Dai}},\ and\ \bibinfo {author} {\bibfnamefont {Z.}~\bibnamefont {Fang}},\
  }\bibfield  {title} {\bibinfo {title} {Dirac semimetal and topological phase
  transitions in ${A}_{3}$bi ($a=\text{Na}$, k, rb)},\ }\href
  {https://doi.org/10.1103/PhysRevB.85.195320} {\bibfield  {journal} {\bibinfo
  {journal} {Phys. Rev. B}\ }\textbf {\bibinfo {volume} {85}},\ \bibinfo
  {pages} {195320} (\bibinfo {year} {2012})}\BibitemShut {NoStop}%
\bibitem [{\citenamefont {Liu}\ \emph {et~al.}(2014{\natexlab{a}})\citenamefont
  {Liu}, \citenamefont {Zhou}, \citenamefont {Zhang}, \citenamefont {Wang},
  \citenamefont {Weng}, \citenamefont {Prabhakaran}, \citenamefont {Mo},
  \citenamefont {Shen}, \citenamefont {Fang}, \citenamefont {Dai} \emph
  {et~al.}}]{Liu2021}%
  \BibitemOpen
  \bibfield  {author} {\bibinfo {author} {\bibfnamefont {Z.}~\bibnamefont
  {Liu}}, \bibinfo {author} {\bibfnamefont {B.}~\bibnamefont {Zhou}}, \bibinfo
  {author} {\bibfnamefont {Y.}~\bibnamefont {Zhang}}, \bibinfo {author}
  {\bibfnamefont {Z.}~\bibnamefont {Wang}}, \bibinfo {author} {\bibfnamefont
  {H.}~\bibnamefont {Weng}}, \bibinfo {author} {\bibfnamefont {D.}~\bibnamefont
  {Prabhakaran}}, \bibinfo {author} {\bibfnamefont {S.-K.}\ \bibnamefont {Mo}},
  \bibinfo {author} {\bibfnamefont {Z.}~\bibnamefont {Shen}}, \bibinfo {author}
  {\bibfnamefont {Z.}~\bibnamefont {Fang}}, \bibinfo {author} {\bibfnamefont
  {X.}~\bibnamefont {Dai}}, \emph {et~al.},\ }\bibfield  {title} {\bibinfo
  {title} {Discovery of a three-dimensional topological dirac semimetal,
  na3bi},\ }\href@noop {} {\bibfield  {journal} {\bibinfo  {journal} {Science}\
  }\textbf {\bibinfo {volume} {343}},\ \bibinfo {pages} {864} (\bibinfo {year}
  {2014}{\natexlab{a}})}\BibitemShut {NoStop}%
\bibitem [{\citenamefont {Liu}\ \emph {et~al.}(2014{\natexlab{b}})\citenamefont
  {Liu}, \citenamefont {Jiang}, \citenamefont {Zhou}, \citenamefont {Wang},
  \citenamefont {Zhang}, \citenamefont {Weng}, \citenamefont {Prabhakaran},
  \citenamefont {Mo}, \citenamefont {Peng}, \citenamefont {Dudin} \emph
  {et~al.}}]{Liu2014}%
  \BibitemOpen
  \bibfield  {author} {\bibinfo {author} {\bibfnamefont {Z.}~\bibnamefont
  {Liu}}, \bibinfo {author} {\bibfnamefont {J.}~\bibnamefont {Jiang}}, \bibinfo
  {author} {\bibfnamefont {B.}~\bibnamefont {Zhou}}, \bibinfo {author}
  {\bibfnamefont {Z.}~\bibnamefont {Wang}}, \bibinfo {author} {\bibfnamefont
  {Y.}~\bibnamefont {Zhang}}, \bibinfo {author} {\bibfnamefont
  {H.}~\bibnamefont {Weng}}, \bibinfo {author} {\bibfnamefont {D.}~\bibnamefont
  {Prabhakaran}}, \bibinfo {author} {\bibfnamefont {S.~K.}\ \bibnamefont {Mo}},
  \bibinfo {author} {\bibfnamefont {H.}~\bibnamefont {Peng}}, \bibinfo {author}
  {\bibfnamefont {P.}~\bibnamefont {Dudin}}, \emph {et~al.},\ }\bibfield
  {title} {\bibinfo {title} {A stable three-dimensional topological dirac
  semimetal cd 3 as 2},\ }\href@noop {} {\bibfield  {journal} {\bibinfo
  {journal} {Nature materials}\ }\textbf {\bibinfo {volume} {13}},\ \bibinfo
  {pages} {677} (\bibinfo {year} {2014}{\natexlab{b}})}\BibitemShut {NoStop}%
\bibitem [{\citenamefont {Wan}\ \emph {et~al.}(2018)\citenamefont {Wan},
  \citenamefont {Schindler}, \citenamefont {Wang}, \citenamefont {Wu},
  \citenamefont {Wan}, \citenamefont {Neupert},\ and\ \citenamefont
  {Lu}}]{Wan2018}%
  \BibitemOpen
  \bibfield  {author} {\bibinfo {author} {\bibfnamefont {B.}~\bibnamefont
  {Wan}}, \bibinfo {author} {\bibfnamefont {F.}~\bibnamefont {Schindler}},
  \bibinfo {author} {\bibfnamefont {K.}~\bibnamefont {Wang}}, \bibinfo {author}
  {\bibfnamefont {K.}~\bibnamefont {Wu}}, \bibinfo {author} {\bibfnamefont
  {X.}~\bibnamefont {Wan}}, \bibinfo {author} {\bibfnamefont {T.}~\bibnamefont
  {Neupert}},\ and\ \bibinfo {author} {\bibfnamefont {H.-Z.}\ \bibnamefont
  {Lu}},\ }\bibfield  {title} {\bibinfo {title} {Theory for the negative
  longitudinal magnetoresistance in the quantum limit of kramers weyl
  semimetals},\ }\href@noop {} {\bibfield  {journal} {\bibinfo  {journal}
  {Journal of Physics: Condensed Matter}\ }\textbf {\bibinfo {volume} {30}},\
  \bibinfo {pages} {505501} (\bibinfo {year} {2018})}\BibitemShut {NoStop}%
\bibitem [{\citenamefont {Li}\ \emph {et~al.}(2021)\citenamefont {Li},
  \citenamefont {Lv}, \citenamefont {Fang}, \citenamefont {Guo}, \citenamefont
  {Wu}, \citenamefont {Wu}, \citenamefont {Shen}, \citenamefont {Nie},
  \citenamefont {Petaccia}, \citenamefont {Cao} \emph {et~al.}}]{li2021charge}%
  \BibitemOpen
  \bibfield  {author} {\bibinfo {author} {\bibfnamefont {P.}~\bibnamefont
  {Li}}, \bibinfo {author} {\bibfnamefont {B.}~\bibnamefont {Lv}}, \bibinfo
  {author} {\bibfnamefont {Y.}~\bibnamefont {Fang}}, \bibinfo {author}
  {\bibfnamefont {W.}~\bibnamefont {Guo}}, \bibinfo {author} {\bibfnamefont
  {Z.}~\bibnamefont {Wu}}, \bibinfo {author} {\bibfnamefont {Y.}~\bibnamefont
  {Wu}}, \bibinfo {author} {\bibfnamefont {D.}~\bibnamefont {Shen}}, \bibinfo
  {author} {\bibfnamefont {Y.}~\bibnamefont {Nie}}, \bibinfo {author}
  {\bibfnamefont {L.}~\bibnamefont {Petaccia}}, \bibinfo {author}
  {\bibfnamefont {C.}~\bibnamefont {Cao}}, \emph {et~al.},\ }\bibfield  {title}
  {\bibinfo {title} {Charge density wave and weak kondo effect in a dirac
  semimetal cesbte},\ }\href@noop {} {\bibfield  {journal} {\bibinfo  {journal}
  {SCIENCE CHINA Physics, Mechanics \& Astronomy}\ }\textbf {\bibinfo {volume}
  {64}},\ \bibinfo {pages} {1} (\bibinfo {year} {2021})}\BibitemShut {NoStop}%
\bibitem [{\citenamefont {Shi}\ \emph {et~al.}(2021)\citenamefont {Shi},
  \citenamefont {Wieder}, \citenamefont {Meyerheim}, \citenamefont {Sun},
  \citenamefont {Zhang}, \citenamefont {Li}, \citenamefont {Shen},
  \citenamefont {Qi}, \citenamefont {Yang}, \citenamefont {Jena} \emph
  {et~al.}}]{shi2021charge}%
  \BibitemOpen
  \bibfield  {author} {\bibinfo {author} {\bibfnamefont {W.}~\bibnamefont
  {Shi}}, \bibinfo {author} {\bibfnamefont {B.~J.}\ \bibnamefont {Wieder}},
  \bibinfo {author} {\bibfnamefont {H.~L.}\ \bibnamefont {Meyerheim}}, \bibinfo
  {author} {\bibfnamefont {Y.}~\bibnamefont {Sun}}, \bibinfo {author}
  {\bibfnamefont {Y.}~\bibnamefont {Zhang}}, \bibinfo {author} {\bibfnamefont
  {Y.}~\bibnamefont {Li}}, \bibinfo {author} {\bibfnamefont {L.}~\bibnamefont
  {Shen}}, \bibinfo {author} {\bibfnamefont {Y.}~\bibnamefont {Qi}}, \bibinfo
  {author} {\bibfnamefont {L.}~\bibnamefont {Yang}}, \bibinfo {author}
  {\bibfnamefont {J.}~\bibnamefont {Jena}}, \emph {et~al.},\ }\bibfield
  {title} {\bibinfo {title} {A charge-density-wave topological semimetal},\
  }\href@noop {} {\bibfield  {journal} {\bibinfo  {journal} {Nature Physics}\
  }\textbf {\bibinfo {volume} {17}},\ \bibinfo {pages} {381} (\bibinfo {year}
  {2021})}\BibitemShut {NoStop}%
\bibitem [{\citenamefont {Zhang}\ \emph {et~al.}(2017)\citenamefont {Zhang},
  \citenamefont {Schindler}, \citenamefont {Liu}, \citenamefont {Chang},
  \citenamefont {Xu}, \citenamefont {Chang}, \citenamefont {Hua}, \citenamefont
  {Jiang}, \citenamefont {Yuan}, \citenamefont {Sun} \emph
  {et~al.}}]{Zhang2017}%
  \BibitemOpen
  \bibfield  {author} {\bibinfo {author} {\bibfnamefont {C.-L.}\ \bibnamefont
  {Zhang}}, \bibinfo {author} {\bibfnamefont {F.}~\bibnamefont {Schindler}},
  \bibinfo {author} {\bibfnamefont {H.}~\bibnamefont {Liu}}, \bibinfo {author}
  {\bibfnamefont {T.-R.}\ \bibnamefont {Chang}}, \bibinfo {author}
  {\bibfnamefont {S.-Y.}\ \bibnamefont {Xu}}, \bibinfo {author} {\bibfnamefont
  {G.}~\bibnamefont {Chang}}, \bibinfo {author} {\bibfnamefont
  {W.}~\bibnamefont {Hua}}, \bibinfo {author} {\bibfnamefont {H.}~\bibnamefont
  {Jiang}}, \bibinfo {author} {\bibfnamefont {Z.}~\bibnamefont {Yuan}},
  \bibinfo {author} {\bibfnamefont {J.}~\bibnamefont {Sun}}, \emph {et~al.},\
  }\bibfield  {title} {\bibinfo {title} {Ultraquantum magnetoresistance in the
  kramers-weyl semimetal candidate $\beta$- ag 2 se},\ }\href@noop {}
  {\bibfield  {journal} {\bibinfo  {journal} {Physical Review B}\ }\textbf
  {\bibinfo {volume} {96}},\ \bibinfo {pages} {165148} (\bibinfo {year}
  {2017})}\BibitemShut {NoStop}%
\bibitem [{\citenamefont {Chang}\ \emph {et~al.}(2018)\citenamefont {Chang},
  \citenamefont {Wieder}, \citenamefont {Schindler}, \citenamefont {Sanchez},
  \citenamefont {Belopolski}, \citenamefont {Huang}, \citenamefont {Singh},
  \citenamefont {Wu}, \citenamefont {Chang}, \citenamefont {Neupert} \emph
  {et~al.}}]{Chang2018}%
  \BibitemOpen
  \bibfield  {author} {\bibinfo {author} {\bibfnamefont {G.}~\bibnamefont
  {Chang}}, \bibinfo {author} {\bibfnamefont {B.~J.}\ \bibnamefont {Wieder}},
  \bibinfo {author} {\bibfnamefont {F.}~\bibnamefont {Schindler}}, \bibinfo
  {author} {\bibfnamefont {D.~S.}\ \bibnamefont {Sanchez}}, \bibinfo {author}
  {\bibfnamefont {I.}~\bibnamefont {Belopolski}}, \bibinfo {author}
  {\bibfnamefont {S.-M.}\ \bibnamefont {Huang}}, \bibinfo {author}
  {\bibfnamefont {B.}~\bibnamefont {Singh}}, \bibinfo {author} {\bibfnamefont
  {D.}~\bibnamefont {Wu}}, \bibinfo {author} {\bibfnamefont {T.-R.}\
  \bibnamefont {Chang}}, \bibinfo {author} {\bibfnamefont {T.}~\bibnamefont
  {Neupert}}, \emph {et~al.},\ }\bibfield  {title} {\bibinfo {title}
  {Topological quantum properties of chiral crystals},\ }\href@noop {}
  {\bibfield  {journal} {\bibinfo  {journal} {Nature materials}\ }\textbf
  {\bibinfo {volume} {17}},\ \bibinfo {pages} {978} (\bibinfo {year}
  {2018})}\BibitemShut {NoStop}%
\bibitem [{\citenamefont {Chen}\ \emph {et~al.}(2021)\citenamefont {Chen},
  \citenamefont {Wu}, \citenamefont {Zhu}, \citenamefont {Gong}, \citenamefont
  {Gao}, \citenamefont {Yang},\ and\ \citenamefont {Zhang}}]{chen2021chiral}%
  \BibitemOpen
  \bibfield  {author} {\bibinfo {author} {\bibfnamefont {H.}~\bibnamefont
  {Chen}}, \bibinfo {author} {\bibfnamefont {W.}~\bibnamefont {Wu}}, \bibinfo
  {author} {\bibfnamefont {J.}~\bibnamefont {Zhu}}, \bibinfo {author}
  {\bibfnamefont {W.}~\bibnamefont {Gong}}, \bibinfo {author} {\bibfnamefont
  {W.}~\bibnamefont {Gao}}, \bibinfo {author} {\bibfnamefont {S.~A.}\
  \bibnamefont {Yang}},\ and\ \bibinfo {author} {\bibfnamefont
  {L.}~\bibnamefont {Zhang}},\ }\bibfield  {title} {\bibinfo {title} {Chiral
  phonons in chiral materials},\ }\href@noop {} {\bibfield  {journal} {\bibinfo
   {journal} {arXiv preprint arXiv:2109.08872}\ } (\bibinfo {year}
  {2021})}\BibitemShut {NoStop}%
\bibitem [{\citenamefont {Portigal}\ and\ \citenamefont
  {Burstein}(1968)}]{PhysRev.170.673}%
  \BibitemOpen
  \bibfield  {author} {\bibinfo {author} {\bibfnamefont {D.~L.}\ \bibnamefont
  {Portigal}}\ and\ \bibinfo {author} {\bibfnamefont {E.}~\bibnamefont
  {Burstein}},\ }\bibfield  {title} {\bibinfo {title} {Acoustical activity and
  other first-order spatial dispersion effects in crystals},\ }\href
  {https://doi.org/10.1103/PhysRev.170.673} {\bibfield  {journal} {\bibinfo
  {journal} {Phys. Rev.}\ }\textbf {\bibinfo {volume} {170}},\ \bibinfo {pages}
  {673} (\bibinfo {year} {1968})}\BibitemShut {NoStop}%
\bibitem [{\citenamefont {Frenzel}\ \emph {et~al.}(2019)\citenamefont
  {Frenzel}, \citenamefont {K{\"o}pfler}, \citenamefont {Jung}, \citenamefont
  {Kadic},\ and\ \citenamefont {Wegener}}]{frenzel2019ultrasound}%
  \BibitemOpen
  \bibfield  {author} {\bibinfo {author} {\bibfnamefont {T.}~\bibnamefont
  {Frenzel}}, \bibinfo {author} {\bibfnamefont {J.}~\bibnamefont
  {K{\"o}pfler}}, \bibinfo {author} {\bibfnamefont {E.}~\bibnamefont {Jung}},
  \bibinfo {author} {\bibfnamefont {M.}~\bibnamefont {Kadic}},\ and\ \bibinfo
  {author} {\bibfnamefont {M.}~\bibnamefont {Wegener}},\ }\bibfield  {title}
  {\bibinfo {title} {Ultrasound experiments on acoustical activity in chiral
  mechanical metamaterials},\ }\href@noop {} {\bibfield  {journal} {\bibinfo
  {journal} {Nature communications}\ }\textbf {\bibinfo {volume} {10}},\
  \bibinfo {pages} {1} (\bibinfo {year} {2019})}\BibitemShut {NoStop}%
\bibitem [{\citenamefont {L{\"u}thi}(2007)}]{luthi2007physical}%
  \BibitemOpen
  \bibfield  {author} {\bibinfo {author} {\bibfnamefont {B.}~\bibnamefont
  {L{\"u}thi}},\ }\href@noop {} {\emph {\bibinfo {title} {Physical acoustics in
  the solid state}}},\ Vol.\ \bibinfo {volume} {148}\ (\bibinfo  {publisher}
  {Springer Science \& Business Media},\ \bibinfo {year} {2007})\BibitemShut
  {NoStop}%
\bibitem [{\citenamefont {Zvezdin}\ and\ \citenamefont
  {Kotov}(1997)}]{zvezdin1997modern}%
  \BibitemOpen
  \bibfield  {author} {\bibinfo {author} {\bibfnamefont {A.~K.}\ \bibnamefont
  {Zvezdin}}\ and\ \bibinfo {author} {\bibfnamefont {V.~A.}\ \bibnamefont
  {Kotov}},\ }\href@noop {} {\emph {\bibinfo {title} {Modern magnetooptics and
  magnetooptical materials}}}\ (\bibinfo  {publisher} {CRC Press},\ \bibinfo
  {year} {1997})\BibitemShut {NoStop}%
\bibitem [{\citenamefont {Lee}\ \emph {et~al.}(1999)\citenamefont {Lee},
  \citenamefont {Haard}, \citenamefont {Halperin},\ and\ \citenamefont
  {Sauls}}]{lee1999discovery}%
  \BibitemOpen
  \bibfield  {author} {\bibinfo {author} {\bibfnamefont {Y.}~\bibnamefont
  {Lee}}, \bibinfo {author} {\bibfnamefont {T.}~\bibnamefont {Haard}}, \bibinfo
  {author} {\bibfnamefont {W.~P.}\ \bibnamefont {Halperin}},\ and\ \bibinfo
  {author} {\bibfnamefont {J.~A.}\ \bibnamefont {Sauls}},\ }\bibfield  {title}
  {\bibinfo {title} {Discovery of the acoustic faraday effect in superfluid 3
  he-b},\ }\href@noop {} {\bibfield  {journal} {\bibinfo  {journal} {Nature}\
  }\textbf {\bibinfo {volume} {400}},\ \bibinfo {pages} {431} (\bibinfo {year}
  {1999})}\BibitemShut {NoStop}%
\bibitem [{\citenamefont {Sauls}\ \emph {et~al.}(2000)\citenamefont {Sauls},
  \citenamefont {Lee}, \citenamefont {Haard},\ and\ \citenamefont
  {Halperin}}]{sauls2000magneto}%
  \BibitemOpen
  \bibfield  {author} {\bibinfo {author} {\bibfnamefont {J.}~\bibnamefont
  {Sauls}}, \bibinfo {author} {\bibfnamefont {Y.}~\bibnamefont {Lee}}, \bibinfo
  {author} {\bibfnamefont {T.}~\bibnamefont {Haard}},\ and\ \bibinfo {author}
  {\bibfnamefont {W.}~\bibnamefont {Halperin}},\ }\bibfield  {title} {\bibinfo
  {title} {Magneto-acoustic rotation of transverse waves in 3he--b},\
  }\href@noop {} {\bibfield  {journal} {\bibinfo  {journal} {Physica B:
  Condensed Matter}\ }\textbf {\bibinfo {volume} {284}},\ \bibinfo {pages}
  {267} (\bibinfo {year} {2000})}\BibitemShut {NoStop}%
\bibitem [{\citenamefont {Einstein}\ and\ \citenamefont
  {De~Haas}(1915)}]{Einstein1915}%
  \BibitemOpen
  \bibfield  {author} {\bibinfo {author} {\bibfnamefont {A.}~\bibnamefont
  {Einstein}}\ and\ \bibinfo {author} {\bibfnamefont {W.}~\bibnamefont
  {De~Haas}},\ }\bibfield  {title} {\bibinfo {title} {Experimental proof of the
  existence of amp{\`e}re’s molecular currents},\ }\href@noop {} {\bibfield
  {journal} {\bibinfo  {journal} {Proc. KNAW}\ }\textbf {\bibinfo {volume}
  {181}},\ \bibinfo {pages} {696} (\bibinfo {year} {1915})}\BibitemShut
  {NoStop}%
\bibitem [{\citenamefont {Zhang}\ and\ \citenamefont {Niu}(2014)}]{Zhang2014}%
  \BibitemOpen
  \bibfield  {author} {\bibinfo {author} {\bibfnamefont {L.}~\bibnamefont
  {Zhang}}\ and\ \bibinfo {author} {\bibfnamefont {Q.}~\bibnamefont {Niu}},\
  }\bibfield  {title} {\bibinfo {title} {Angular momentum of phonons and the
  einstein--de haas effect},\ }\href@noop {} {\bibfield  {journal} {\bibinfo
  {journal} {Physical Review Letters}\ }\textbf {\bibinfo {volume} {112}},\
  \bibinfo {pages} {085503} (\bibinfo {year} {2014})}\BibitemShut {NoStop}%
\bibitem [{TAC()}]{TACC}%
  \BibitemOpen
  \href@noop {} {\bibinfo {title} {Texas advanced computing center}},\ \bibinfo
  {howpublished} {\url{https://www.tacc.utexas.edu/}}\BibitemShut {NoStop}%
\end{thebibliography}%

\onecolumngrid 
\appendix

\section{Charged Harmonic Oscillator in Uniform Magnetic Field\label{app:chargedHO}}
To estimate the magnitude of direct coupling of magnetic field on lattice (ions) vibration, here we consider a simplest 3D harmonic oscillator carrying $Ze$ charges. By minimal coupling, its Hamiltonian reads,
    \begin{equation}\begin{aligned}
        \hat{H}^{ph}
        =&\frac{1}{2M}[(\hat{P}_{x}-ZeA_{x})^{2}+(\hat{P}_{y}-ZeA_{y})^{2}+\hat{P}_{z}^{2}]
            +\frac{1}{2}M\omega_{T}^{2}(\hat{X}_{x}^{2}+\hat{X}_{y}^{2})+\frac{1}{2}M\omega_{z}^{2}\hat{X}_{z}^{2}\\
        =&\left[\frac{1}{2M}(\hat{P}_{x}^{2}+\hat{P}_{y}^{2})+\frac{1}{2}M\left(\omega_{T}^{2}+(\frac{ZeB}{2M})^{2}\right)(\hat{X}_{x}^{2}+\hat{X}_{y}^{2})\right]
        +\left(\frac{1}{2M}\hat{P}_{z}^{2}+\frac{1}{2}M\omega_{z}^{2}\hat{X}_{z}^{2}\right)
        -\frac{ZeB}{2M}\hat{L}_{z}\\
        =&\left[\frac{1}{2M}(\hat{P}_{x}^{2}+\hat{P}_{y}^{2})+\frac{1}{2}M\tilde{\omega}_{T}^{2}(\hat{X}_{x}^{2}+\hat{X}_{y}^{2})\right]
        +\left(\frac{1}{2M}\hat{P}_{z}^{2}+\frac{1}{2}M\omega_{z}^{2}\hat{X}_{z}^{2}\right)
        -\frac{Zm_{e}}{M\hbar}\mu_{B}B\hat{L}_{z},\\
    \tilde{\omega}_{T}^{2}
        =&\omega_{T}^{2}+\left(\frac{Zm_{e}}{M\hbar}\mu_{B}B\right)^{2},
    \end{aligned}\end{equation}
where $M$ is the mass, and $\omega_{z}$ and $\omega_{T}$ are eigen-energyies of $z$-axis and $xy$-plane (given rotation symmetry along z-axis). Comparing to the unperturbed Hamiltonian $\hat{H}^{ph}(\bm{A}=\bm{0})$, two in-plane effects are introduced: the degenerate eigen-frequencies of states in $xy$-plane shifting from $\omega_{T}$ to $\tilde{\omega}_{T}$, and the splitting $\Delta\omega_{T}=2(Zm_{e}\mu_{B}B/M\hbar^{2})\hat{L}_{z}$ between chiral modes with different orbital angular momenta. Now we can do a back-of-the-envelope calculation: assuming that the frequency of a certain phonon mode in realistic materials is around $\omega_{T}=1THz$ (speed $v_{T}^{ph}=2\times 10^{3}m/s$ and wave vector $q_{z}=0.5\times 10^{9}m^{-1}$, smaller than 1/10 scale of the first BZ), the mass ratio of the electron to the ion $m_{e}/M\sim 10^{-6}$, the effective charge of ion $Z\sim 10$, and $\mu_{B}B/\hbar\approx 1THz$ when $B=10T$. Taking the unperturbed frequency $\omega_{T}$ as a reference, the order of magnitude of shifting and splitting are around $|\tilde{\omega}_{T}-\omega_{T}|/\omega_{T}\approx 10^{-10}$ and $\Delta\omega_{T}/\omega_{T}\approx 10^{-5}$ given $|L_{z}|=\hbar$. Therefore, it is sensible to ignore both two effects and treat phonons as neutral particles when we investigate electron-phonon interaction.

\section{Selection Rules in Symmetric Gauge\label{app:selectionRule}}
In this section, we specify that, in the quantum limit, indices $n$ and $m$ are conserved separately and then derive the selection rules in electron-phonon interaction. Let us start with the orbital angular momentum of the spinless 2D electron gas. When a uniform magnetic field with symmetric gauge $\bm{A}=(-y,x,0)B/2$ applied, the Hamiltonian reads,
    \begin{equation}\begin{aligned}
        \hat{H}^{2D}
        =\frac{1}{2m}[(-i\hbar\partial_{x}-yeB/2)^{2}+(-i\hbar\partial_{y}+xeB/2)^{2}]
        =\frac{1}{2m}(\hat{\pi}_{x}^{2}+\hat{\pi}_{y}^{2}),
    \end{aligned}\end{equation}
where $\hat{\pi}_{x,y}$ are covariant momenta. Based on the guiding-center coordinates
$\hat{r}_{x}=\hat{x}+\hat{\pi}_{y}/eB$ and $\hat{r}_{y}=\hat{y}-\hat{\pi}_{x}/eB$, we can define two annihilation operators,
    \begin{equation}
        \hat{a}=\frac{l_{B}}{\sqrt{2}\hbar}(\hat{\pi}_{x}+i\hat{\pi}_{y}),~
        \hat{b}=\frac{l_{B}}{\sqrt{2}\hbar}(\hat{r}_{x}-i\hat{r}_{y})
    \end{equation}
with magnetic length $l_{B}=\sqrt{\hbar/eB}$ and commutation relations 
$[\hat{a},\hat{a}^{\dagger}]=[\hat{b},\hat{b}^{\dagger}]=1$,
$[\hat{a},\hat{b}]=[\hat{a}^{\dagger},\hat{b}]=0$. 
Then all Landau wave functions are able to be constructed as $|n,m\rangle=(n!m!)^{-1/2}(\hat{a}^{\dagger})^{n}(\hat{b}^{\dagger})^{m}|0\rangle$,
where $n,m$ are Landau level index and sub-index, respectively. Then the Hamiltonian can be rewritten as $\hat{H}=(\hat{a}^{\dagger}\hat{a}+1/2)\hbar\omega_{c}$ (cyclotron frequency $\omega_{c}=eB/m$), which is only dependent on Landau level index and all Landau levels are highly degenerate with different $m$ quantum number. 
The continuous rotation symmetry gives us a conserved quantity, canonical orbital angular momentum
    \begin{equation}\begin{aligned}
        \hat{L}_{z}
        &=-i\hbar(x\partial_{y}-y\partial_{x})
        =\frac{B}{2}(\hat{r}_{x}^{2}+\hat{r}_{y}^{2})-\frac{1}{2eB}(\hat{\pi}_{x}^{2}+\hat{\pi}_{y}^{2})
        =(\hat{b}^{\dagger}\hat{b}-\hat{a}^{\dagger}\hat{a})\hbar\equiv\hat{L}_{z}^{(m)}+\hat{L}_{z}^{(n)},
    \end{aligned}\end{equation}
where $\hat{L}_{z}^{(n)}=-\hat{a}^{\dagger}\hat{a}\hbar$, $\hat{L}_{z}^{(m)}=\hat{b}^{\dagger}\hat{b}\hbar$ and $[\hat{L}_{z}^{(n)},\hat{L}_{z}^{(m)}]=0$. Now we know, a state $|n,m\rangle$ carries orbital angular momentum $L_{z}=(m-n)\hbar$. 

For a generic continuum model of 3D spinful electron under magnetic field $B_{z}\hat{\bm{z}}$, the $xy$-plane orbital motions are still described on basis of Landau wavefunctions $|n,m\rangle$, and only the total angular momentum $(s_{z}+m-n)\hbar$ is conserved. 
Since Hamiltonian $\mathcal{H}^{\alpha\beta}_{k_{z}}(-i\partial_{x}+eA_{x}/\hbar c,-i\partial_{y}+eA_{y}/\hbar c)$ is quantized to $\hat{\mathcal{H}}^{\alpha\beta}_{k_{z}}((\hat{a}^{\dagger}+\hat{a})/\sqrt{2}l_{B},(\hat{a}^{\dagger}-\hat{a})/i\sqrt{2}l_{B})$, all SOC terms are only related to $L_{z}^{(n)}$ and irrelevant to $L_{z}^{(m)}$. Therefore, the conservation of angular momenta are actually separated into two parts
    \begin{equation}
        s_{z}+n = s_{z}'+n',~m = m'.
    \end{equation}
As will be discussed in the next section, the electron-phonon interaction we obtained only involves the spin flip and  happens within Landau bands with the same indices $n=n'$, so the selection rules arrive at a quite demanding form,
    \begin{equation}
        s_{z} = s_{z}'+l_{z,\lambda}^{ph},~m = m',
    \end{equation}
where $l_{z,\lambda}^{ph}$ is the orbital angular momentum of certain phonon branch with polarization $\lambda$.

\section{Electron-Phonon Interaction}
\subsection{Bir-Pikus Formalism\label{app:B-1}}

The lattice vibration in the continuum limit is a \textit{time-dependent local} strain $\hat{\bm{\varepsilon}}(\bm{r},t)$. For the small amplitude of ions' displacement $\hat{u}_{\lambda }(\bm{r})=N^{-1/2}\sum_{\bm{q}}\hat{X}_{\lambda\bm{q}}e^{i\bm{q}\cdot\bm{r}}$ compared with the scale of the unit cell, the quantized strain operator in terms of phonon creation and annihilation operators is 
    \begin{equation}\begin{aligned}
        \hat{\varepsilon}_{ij}(\bm{r})
        =&\frac{1}{\sqrt{N}}\sum_{\bm{q}}iq_{j}\hat{X}_{i\bm{q}}e^{i\bm{q}\cdot\bm{r}}
        =\frac{1}{\sqrt{N}}\sum_{\bm{q}}iq_{j}\xi_{i\bm{q}}(\hat{b}_{i\bm{q}}+\hat{b}_{i,-\bm{q}}^{\dagger})e^{i\bm{q}\cdot\bm{r}},
    \end{aligned}\end{equation}
with $\xi_{\lambda\bm{q}}=\sqrt{\hbar/2M\omega_{\lambda\bm{q}}}$, and it is naturally to treat it by perturbation theory. 
However, a generic deformed lattice potential $V^{\varepsilon}(\bm{r})$ do not have the same periodicity as the original one $V^{0}(\bm{r})$ and the regular perturbation theory is not justified any more since the wave function of the perturbed Hamiltonian is always expressed as a superposition of wave functions of the unperturbed $\hat{H}^{0}$ satisfying the same boundary conditions. 
The same difficulty shows up when Bir and Pikus tried to tame the effect of a homogeneous strain. Therefore, we develop their formalism to deal with the effect of lattice vibration and transform the coordinates to make the periodicity in the new coordinate system coincide with the un-strained situation in the old coordinate system. Up to linear order of strain, this is done by putting
\begin{equation}
    r_{i}'=r_{i}+\hat{\varepsilon}_{ij}r_{j},~
    \hat{p}_{i}'=-i\hbar\frac{\partial}{\partial r_{i}'}=\hat{p}_{i}-\hat{\varepsilon}_{ij}\hat{p}_{j},
\end{equation}
where $\bm{r}$, $\bm{p}$ are electron's coordinate and momentum, and the transformation between the reciprocal vectors is
\begin{equation}
    k_{i}'=k_{i}-\hat{\varepsilon}_{ij}k_{j}.
\end{equation}
Correspondingly, the Bloch function in the deformed system becomes,
\begin{equation}
    e^{i\bm{k}'\cdot\bm{r}'}u_{n\bm{k}'}(\bm{r}')    
    =e^{i\bm{k}\cdot\bm{r}}u_{n \bm{k}'} ((1+\hat{\bm{\varepsilon}})\bm{r})
    \equiv e^{i\bm{k}\cdot\bm{r}}u_{n\bm{k}}'(\bm{r})
\end{equation}
having the same phase factor as in undeformed system. 
By this way, we are safe to use perturbation theory and their difference can be expanded in terms of $\hat{\varepsilon}_{ij}$,
\begin{equation}\begin{split}
  \hat{H}^{0} 
  =& \frac{\hat{\bm{p}}^2}{2 m_0} + V^{0}(\bm{r}) + \frac{\hbar}{4 m_0^2 c^2} (\hat{\bm{\sigma}} \times \nabla V^{0}) \cdot \hat{\bm{p}},\\
  \hat{H}^{ep} 
  =& \hat{H}^{\varepsilon}(\bm{r}',\hat{\bm{p}}')-\hat{H}^{0}(\bm{r},\hat{\bm{p}}),
\end{split}\end{equation}
so the strained electron Hamiltonian in the original coordinates transforms into the deformed coordinates:
\begin{equation}\begin{split}
  \hat{H}^{\varepsilon}(\bm{r}',\hat{\bm{p}}')
  =& \frac{\hat{\bm{p}}'^{2}}{2m_{0}} + V^{\varepsilon}(\bm{r}')
    +\frac{\hbar}{4m_0^{2}c^{2}}(\hat{\bm{\sigma}}\times\nabla' V^{\varepsilon}(\bm{r}'))\cdot\hat{\bm{p}}'\\
  =& \frac{[(1-\hat{\bm{\varepsilon}})\hat{\bm{p}}]^{2}}{2 m_{0}}+V^{\varepsilon}[(1+\hat{\bm{\varepsilon}})\bm{r}]
  + \frac{\hbar}{4m_0^{2}c^{2}}[\hat{\bm{\sigma}}\times  ((1-\hat{\bm{\varepsilon}})\nabla) V^{\varepsilon}((1+\hat{\bm{\varepsilon}})\bm{r})]\cdot (1-\hat{\bm{\varepsilon}})\hat{\bm{p}}\\
  =& \hat{H}^{0}(\bm{r},\hat{\bm{p}})
   - \frac{1}{m_0}\hat{p}_{i}\hat{\varepsilon}_{ij}\hat{p}_{j} 
   + V_{ij}\hat{\varepsilon}_{ij}
   - \frac{\hbar}{4m_0^{2}c^{2}}\epsilon_{ijk}\hat{\sigma}_{i}[(\partial_{j}V^{0})\hat{\varepsilon}_{kt} \hat{p}_t
  + (\hat{\varepsilon}_{jl}\partial_{l}V^{0})\hat{p}_k 
  - (\partial_{j}V_{lm}\hat{\varepsilon}_{lm}) \hat{p}_k]
\end{split}\end{equation}
where $V_{ij}\equiv\lim_{\varepsilon\rightarrow 0}\{V^{\varepsilon}[(1+\hat{\varepsilon})\bm{r}]-V^{0}(\bm{r})\}/\hat{\varepsilon}_{ij}$, and we used commutation relation     
  \begin{equation}
  [\hat{p}_{i},\hat{\varepsilon}_{jk}]
  =\frac{1}{\sqrt{N}}\sum_{\bm{q}}iq_{k}\xi_{j\bm{q}}(\hat{b}_{j\bm{q}}+\hat{b}^{\dagger}_{j,-\bm{q}})[\hat{p}_{i},e^{i\bm{q}\cdot\bm{R}}]
  =\frac{1}{\sqrt{N}}\sum_{\bm{q}}iq_{k}\xi_{j\bm{q}}(\hat{b}_{j\bm{q}}+\hat{b}^{\dagger}_{j,-\bm{q}})(\hbar q_{i}e^{i\bm{q}\cdot\bm{R}})
  =0+\mathcal{O}(q^{2}),
  \end{equation}
then,
    \begin{equation}\begin{split}
   &((1-\hat{\bm{\varepsilon}})\nabla)  V^{\varepsilon}((1+\hat{\bm{\varepsilon}})\bm{r})\\
  =&(\partial_{i}-\hat{\varepsilon}_{ij}\partial_j) (\hat{V}_0 + V_{lm}\hat{\varepsilon}_{lm}) \hat{e}_i 
  =(\partial_{i}V^{0} -\hat{\varepsilon}_{ij}
  \partial_{j}V^{0}+\partial_{i}{V}_{lm} \hat{\varepsilon}_{lm}) \hat{e}_i,\\
   &[\hat{\bm{\sigma}}\times((1-\hat{\bm{\varepsilon}})\nabla) V^{\varepsilon}((1+\hat{\bm{\varepsilon}})\bm{r})]\cdot (1-\hat{\bm{\varepsilon}})\hat{\bm{p}}\\
  =& \epsilon_{ijk}\hat{\sigma}_{i}(\partial_{j}V^{0} - \hat{\varepsilon}_{jl} \partial_l V_0
  + \partial_{j}V_{lm}\hat{\varepsilon}_{lm})(\hat{p}_k - \hat{\varepsilon}_{kt}\hat{p}_t)\\
  =& \epsilon_{ijk}\hat{\sigma}_{i}[(\partial_{j}V^{0}) \hat{p}_{k}-(\partial_{j}V^{0})\hat{\varepsilon}_{kt} \hat{p}_t
   - (\hat{\varepsilon}_{jl}\partial_{l}V^{0})\hat{p}_k + (\partial_{j}V_{lm}\hat{\varepsilon}_{lm}) \hat{p}_k]\\
  =&\hat{\bm{\sigma}}\times\nabla V^{0}\cdot\hat{\bm{p}}
    -\hat{\bm{\sigma}}\times\nabla V^{0}\cdot(\hat{\bm{\varepsilon}}\hat{\bm{p}})
   - \hat{\bm{\sigma}}\times(\hat{\bm{\varepsilon}}\nabla  V^{0})\cdot\hat{\bm{p}}
    +\hat{\bm{\sigma}}\times\nabla (\hat{\bm{\varepsilon}}\bm{V})\cdot\hat{\bm{p}},\\
    \end{split}\end{equation}   
where we denote $\hat{\bm{\varepsilon}}\bm{V}\equiv \hat{\varepsilon}_{lm}V_{lm}$ following the original notation by Bir and Pikus.
Then we have,
    \begin{equation}\begin{split}
     \hat{H}^{ep}
  = & \bm{V}\hat{\bm{\varepsilon}} -
    \frac{1}{m_0} \hat{\bm{p}}\hat{\bm{\varepsilon}}\hat{\bm{p}}
   + \frac{\hbar}{4 m_0^2 c^2}[\hat{\bm{\sigma}}\times\nabla (\hat{\bm{\varepsilon}}\bm{V})\cdot\hat{\bm{p}}
   - \hat{\bm{\sigma}} \times \nabla V^{0}\cdot(\hat{\bm{\varepsilon}}\hat{\bm{p}})
  - \hat{\bm{\sigma}}\times(\hat{\bm{\varepsilon}}\nabla  V^{0})\cdot\hat{\bm{p}}]\\
  = &  -\frac{1}{m_0} \hat{\pi}_{i}\hat{\varepsilon}_{ij}\hat{p}_{j}
   - \frac{\hbar}{4 m_0^2 c^2}(\hat{\bm{p}}\times\hat{\bm{\sigma}})_{i}\hat{\varepsilon}_{ij}\partial_{j}V^{0}
   + V_{ij}\hat{\varepsilon}_{ij}
   + \frac{\hbar}{4 m_0^2 c^2}\hat{\bm{\sigma}}\times(\hat{\varepsilon}_{ij}\nabla V_{ij})\cdot\hat{\bm{p}}\\
  = &\hat{\varepsilon}_{ij}\hat{\mathcal{V}}_{ij}+\mathcal{O}(q^{2})\\
    \end{split}\end{equation}
where $\hat{\bm{\pi}} = \hat{\bm{p}} + (\hat{\bm{\sigma}} \times \nabla \hat{V}_0)\hbar/4 m_0 c^2$, and note that  $\hat{\bm{\sigma}}\times(\nabla\hat{\bm{\varepsilon}})\bm{V}\cdot\hat{\bm{p}}$ is in the order of $q^{2}$. Here $\hat{\mathcal{V}}_{ij}$ is a rank-2 tensor operator. 
Then the Schrodinger equation becomes
    \begin{equation}\begin{aligned}
    (\hat{H}^{0}+\hat{H}^{ep})e^{i\bm{k}\cdot\bm{r}}u_{n\bm{k}}'(\bm{r})
    =E_{n\bm{k}}' e^{i\bm{k}\cdot\bm{r}}u_{n\bm{k}}'(\bm{r})
    \Rightarrow
    \hat{H}^{\varepsilon}(\bm{k})u_{n\bm{k}}'(\bm{r})
    =E_{n\bm{k}}'u_{n\bm{k}}'(\bm{r}).
    \end{aligned}\end{equation}
In the end, we arrive at the strained $k\cdot p$ Hamiltonian,
    \begin{equation}\begin{split}\label{eq:EPI}
  \hat{H}^{\varepsilon}({\bm{k}}) 
  = & \hat{H}^{0} + \hat{H}^{0}_{\bm{k}} + \hat{H}^{ep}+\hat{H}^{ep}_{\bm{k}},\\
  \hat{H}^{0}_{\bm{k}}
  =&\frac{\hbar^2 k^2}{2 m_0} +\frac{\hbar}{m_0}\bm{k}\cdot \hat{\bm{\pi}},\\
  \hat{H}^{ep}_{\bm{k}}
  =& \frac{\hbar^{2}\hat{\varepsilon}_{ij}k_{i}k_{j}}{m_0}
    +\frac{\hbar}{m_0}(\hat{\varepsilon}_{ij}+\hat{\varepsilon}_{ji})k_{i}\hat{p}_{j}
   + \frac{\hbar^{2}}{4 m_0^2 c^2}[\hat{\bm{\sigma}}\times\nabla (\hat{\bm{\varepsilon}}\bm{V})\cdot\bm{k}
   - \hat{\bm{\sigma}} \times \nabla V^{0}\cdot(\hat{\bm{\varepsilon}}\bm{k})
    - \hat{\bm{\sigma}}\times(\hat{\bm{\varepsilon}}\nabla  V^{0})\cdot\bm{k}]
    \end{split}\end{equation}
where the first two terms of $\hat{H}^{ep}_{\bm{k}}$ are non-relativistic effects and all others are the contributions from SOC. 
In this work, we only consider electron-phonon interaction up to the zeroth order of $\bm{k}$, so $\hat{H}_{\bm{k}}^{ep}$ is ignored in the following. Note that, if we are interested in Hamiltonian up to first order of $\bm{q}$, treat strain tensor $\bm{\varepsilon}$ as a classical quantity during all the derivations and quantize it at the very end in Eq.(\ref{eq:EPI}) will obtain the same result. 
Because we focus on phonons with well-defined angular momenta along $z$-direction, i.e., non-zero components are $\hat{\varepsilon}_{iz}(i=x,y,z)$, we only need to concern $\hat{\mathcal{V}}_{iz}$ terms and the others have no contributions to electron-phonon interaction. Up to the coefficients, the spherical components are defined as
    \begin{equation}\begin{aligned}
    \hat{\mathcal{V}}_{0}
    =\hat{\mathcal{V}}_{zz},~
    \hat{\mathcal{V}}_{\pm 1}
    =\frac{1}{2}(\hat{\mathcal{V}}_{xz}\pm i\hat{\mathcal{V}}_{yz}),~
    \hat{\mathcal{V}}_{\pm 2}=0.
    \end{aligned}\end{equation}
As followed, we rewrite the electron-phonon interaction in terms of these symmetric components,
    \begin{equation}\begin{aligned}
    \hat{H}^{ep}
    =\hat{\varepsilon}_{zz}\hat{\mathcal{V}}_{zz}+\hat{\varepsilon}_{+,z}\hat{\mathcal{V}}_{-1}+\hat{\varepsilon}_{-,z}\hat{\mathcal{V}}_{+1},
    \end{aligned}\end{equation}
where $\hat{\varepsilon}_{\pm,z}=\hat{\varepsilon}_{x,z}\pm i\hat{\varepsilon}_{y,z}$. 
Now we arrive at the matrix elements of electron-phonon interaction Hamiltonian on chiral phonon basis,
    \begin{equation}\begin{split}
    &\langle\psi_{s_{z}^{\alpha}mk_{z}}|\hat{\varepsilon}_{\pm,z}\hat{\mathcal{V}}_{\mp1}|\psi_{s_{z}^{\beta}m'k_{z}'}\rangle\\
    =&\frac{1}{\sqrt{N}}iq_{z}\xi_{\pm,q_{z}}(\hat{b}_{\pm,q_{z}}+\hat{b}_{\pm,-q_{z}}^{\dagger})
        \langle u_{s_{z}^{\alpha}mk_{z}}|e^{i(k_{z}'+q_{z}-k_{z})z}\hat{\mathcal{V}}_{\mp1}|u_{s_{z}^{\beta}m'k_{z}'}\rangle\delta(m'-m)\\
    =&\frac{1}{\sqrt{N}}iq_{z}\xi_{\pm,q_{z}}(\hat{b}_{\pm,q_{z}}+\hat{b}_{\pm,-q_{z}}^{\dagger})
        \langle u_{s_{z}^{\alpha}mk_{z}+q_{z}}|\hat{\mathcal{V}}_{\mp1}|u_{s_{z}^{\beta}mk_{z}}\rangle
        \delta(k_{z}'+q_{z}-k_{z})\delta(s_{z}^{\alpha}-s_{z}^{\beta}-l_{z,\pm,q_{z}}^{ph})\\
    \equiv&\frac{1}{\sqrt{N}}\mathcal{G}_{mk_{z}q_{z}}^{\pm}(\hat{b}_{\pm,q_{z}}+\hat{b}_{\pm,-q_{z}}^{\dagger}),\\
    \end{split}\end{equation}
where $\hat{\mathcal{V}}$ has experienced Landau quantization procedure, $|u_{s_{z}^{\beta}mk_{z}}\rangle$ is the periodic part of Bloch wave function $|\psi_{s_{z}^{\beta}mk_{z}}\rangle$, and $\hat{H}^{ep}$ is diagonal with respect to $m$ index due to the selection rules. Similarly, for LA phonon we have,
    \begin{equation}\begin{split}
    &\langle \psi_{s_{z}^{\alpha}mk_{z}}|\hat{\varepsilon}_{zz}\hat{\mathcal{V}}_{zz}|\psi_{s_{z}^{\beta}m'k_{z}'}\rangle\\
    =&\frac{1}{\sqrt{N}}iq_{z}\xi_{zq_{z}}(\hat{b}_{zq_{z}}+\hat{b}_{z,-q_{z}}^{\dagger})
        \langle u_{s_{z}^{\alpha}mk_{z}+q_{z}}|\hat{\mathcal{V}}_{zz}|u_{s_{z}^{\beta}mk_{z}}\rangle
         \delta(k_{z}'+q_{z}-k_{z})\delta(s_{z}^{\alpha}-s_{z}^{\beta})\\
    \equiv&\frac{1}{\sqrt{N}}\mathcal{G}_{mk_{z}q_{z}}^{z}(\hat{b}_{z,q_{z}}+\hat{b}_{z,-q_{z}}^{\dagger}).\\
    \end{split}\end{equation}

In realistic materials, it is not practical to calculate these integrals $\langle u_{s_{z}^{\alpha}mk_{z}+q_{z}}|\hat{\mathcal{V}}_{\lambda z}|u_{s_{z}^{\beta}mk_{z}}\rangle(\lambda=\pm,z)$ either analytically or numerically. Thus, in the next section, we parametrize them based on the principles of symmetries.

\subsection{Constrained by Point Group Symmetries\label{app:B-2}}

In certain materials, electron-phonon interaction is constrained by the crystal symmetries. For Na$_{3}$Bi (No.194 space group $P6_{3}/mmc$) at $\Gamma$ point, the point group is $6/mmm$ generated by rotations $C_{3z}$, $C_{2z}$, $C_{110}$ and inversion $P$. For $\beta$-Ag$_{2}$Se (No.19 space group $P2_{1}2_{1}2_{1}$) at $\Gamma$ point, the point group is $222$, generated by $C_{2z}$ and $C_{2x}$.
Here we consider three symmorphic symmetryies of our interest: rotation, inversion and mirror as their combination. 
Starting with the rotational symmetry, on the basis 
$\hat{\Psi}^{\dagger}_{mk_{z}Q} 
= (\hat{c}_{\alpha m k_{z}+Q/2}^{\dagger},\hat{c}_{\beta m k_{z}+Q/2}^{\dagger}, 
\hat{c}_{\alpha m k_{z}-Q/2}^{\dagger}, \hat{c}_{\beta m k_{z}-Q/2}^{\dagger})^T$ (here we assume $s_{z}^{\alpha}>s_{z}^{\beta}$), up to the zeroth order of $k_z$, with
the selection rule $\pm(s_z^{\alpha}-s_z^{\beta}) = l_{z,\mp,Q}^{ph}$. And the angular momentum of chiral phonon is defined as
    \begin{equation}
    \hat{L}_{z,q_{z}}
    =i\hbar(\hat{b}_{x,q_{z}}^{\dagger}\hat{b}_{y,q_{z}}-\hat{b}_{y,q_{z}}^{\dagger}\hat{b}_{x,q_{z}})
    =\hbar(\hat{b}_{-,q_{z}}^{\dagger}\hat{b}_{-,q_{z}}-\hat{b}_{+,q_{z}}^{\dagger}\hat{b}_{+,q_{z}}).
    \end{equation}
where $\hat{b}_{\pm,q_{z}}^{\dagger}=(\hat{b}_{x,q_{z}}^{\dagger}\pm i\hat{b}_{y,q_{z}}^{\dagger})/\sqrt{2}$, $\hat{b}_{\pm,q_{z}}=(\hat{b}_{x,q_{z}}\mp i\hat{b}_{y,q_{z}})/\sqrt{2}$ and correspondingly $\hat{X}_{\pm,q_{z}}=(\hat{X}_{x,q_{z}}\pm i\hat{X}_{y,q_{z}})/\sqrt{2}=\xi_{\pm,q_{z}}(\hat{b}_{\mp,q_{z}}+\hat{b}_{\pm,-q_{z}}^{\dagger})$ given amplitude $\xi_{x,q_{z}}=\xi_{y,q_{z}}=\xi_{\pm,q_{z}}$ due to the rotation symmetry.

Consider a certain phonon mode $\hat{u}_{\lambda q_{z}}(z,t)=\hat{X}_{\lambda q_{z}}(t)e^{iq_{z}z}+\hat{X}_{\lambda,-q_{z}}(t)e^{-iq_{z}z}$, the electron-phonon interaction Hamiltonian for the same $n$ index Landau bands reads
    \begin{equation}\begin{split}
    \hat{\mathcal{H}}^{ep}_{mk_{z}q_{z}} 
    =&\frac{1}{\sqrt{N}}
       \sum_{\lambda}\left[\tau_{+}\mathcal{G}_{mk_{z}q_{z}}^{\lambda}\hat{X}_{\lambda q_{z}}/\xi_{\lambda q_{z}} +h.c.\right]
       \hat{\Psi}_{mk_{z}q_{z}}^{\dagger}\hat{\Psi}_{mk_{z}q_{z}},\\
    \mathcal{G}_{mk_{z}q_{z}}^{\lambda=z}
        =&iq_{z}\xi_{zq_{z}}(g_{z0}\sigma_{0}+g_{z3}\sigma_{3}),\\
    \mathcal{G}_{mk_{z}q_{z}}^{\lambda=\pm}
        =&iq_{z}\xi_{\lambda q_{z}}g_{\pm}\sigma_{\mp},
    \end{split}\end{equation}
where $g_{z0,3}=(g_{z1}\pm g_{z2})/2$, $\epsilon_{\lambda\lambda'}$ the 2D Levi-Civita antisymmetric tensor, the coefficients are complex constants determined by specific materials. 
Note that, in the case of $|\bm{q}|=0$, three acoustic normal modes correspond to global translations of the crystal, of which the displacement amplitude is not defined and such translations will not alter the electronic band structure. 
Therefore, it is assumed that these modes are skipped throughout this article. Note that, here $g_{i}$ are complex number. Considering that $\sigma_{z}$ is the difference of the coupling strengths between LA phonon and electrons in two different bands, we just drop it in the main text.

To make the physical meaning of electron-phonon interaction more clear, we rewrite the above electron-phonon interaction  $\mathcal{G}_{mk_{z}q_{z}}^{\lambda=\pm}$ on the linearly polarized basis, 
    \begin{equation}\begin{split}
    \sum_{\lambda=\pm}\mathcal{G}_{mk_{z}q_{z}}^{\lambda}\hat{X}_{\lambda q_{z}}/\xi_{\lambda q_{z}}
        =&iq_{z}\begin{pmatrix}
                0&g_{-}(\hat{X}_{x q_{z}}-i\hat{X}_{y q_{z}})\\
                g_{+}(\hat{X}_{x q_{z}}+i\hat{X}_{y q_{z}}) & 0
                \end{pmatrix}\\
        =&iq_{z}\begin{pmatrix}
                0&(g_{1}-ig_{2})(\hat{X}_{x q_{z}}-i\hat{X}_{y q_{z}})\\
                 (g_{1}+ig_{2})(\hat{X}_{x q_{z}}+i\hat{X}_{y q_{z}}) & 0
                \end{pmatrix}\\
        \equiv &iq_{z}(g_{1}\hat{\bm{X}}_{Tq_{z}}\cdot\bm{\sigma}_{T} + g_{2}\hat{\bm{X}}_{Tq_{z}}\times\bm{\sigma}_{T})
    \end{split}\end{equation}
where $g_{1}=(g_{-}+g_{+})/2,g_{2}=i(g_{-}-g_{+})/2$, 
$\hat{\bm{X}}_{Tq_{z}}=(\hat{X}_{xq_{x}},\hat{X}_{yq_{z}})$, and $\bm{\sigma}_{T}=(\sigma_{x},\sigma_{y})$. This is exactly two possible interaction between two vector operators: TA phonon normal coordinate and electron spin. 

If SIS is considered, as in the Dirac semimetal case, electron-phonon interaction is supposed to be further constrained by
$\mathcal{P}^{-1}\mathcal{G}_{k_{z}q_z}^{\lambda}\mathcal{P} = \hat{\mathcal{G}}_{-k_{z},-q_z}^{\lambda\dagger}$
with $\mathcal{P}=\sigma_{z}$, then $g_{+}=-g_{-}^{\star}$. However, in Kramers-Weyl semimetal, to emphsize the absence of SIS, we set $g_{+}=0$, i.e., $\mathcal{G}_{mk_{z}q_{z}}^{+}=0$, corresponding to the limit case where left-handed TA phonon is decoupled with low-energy electrons. 
Given additional mirror symmetry $\mathcal{M}_{x}^{-1}\mathcal{G}_{q_z}(\hat{X}_{xq_{z}},\hat{X}_{yq_{z}})\mathcal{M}_{x} = \mathcal{G}_{q_z}(\hat{X}_{xq_{z}},-\hat{X}_{yq_{z}})$ with $\mathcal{M}_{x}=\mathcal{P}\mathcal{C}_{2x}=\sigma_{y}$ further constrained, the $\hat{\bm{X}}_{Tq_{z}}\cdot\bm{\sigma}_{T}$ term is killed and we have
    \begin{equation}\begin{aligned}
    \sum_{\lambda=\pm}\mathcal{G}_{mk_{z}q_{z}}^{\lambda}\hat{X}_{\lambda q_{z}}/\xi_{\lambda q_{z}}
    &= iq_{z}g_{2}\hat{\bm{X}}_{Tq_{z}}\times\bm{\sigma}_{T}.\\
    \end{aligned}\end{equation}

\section{Effective Models in Quantum Limit}

\subsection{Topological Dirac Semimetal\label{app:C-1}}

Near $\Gamma$ point, on the basis 
$\hat{\Psi}_{\bm{k}}^{D\dagger} = \left( \hat{c}_{1/2,\bm{k}}^{\dagger},\hat{c}_{-1/2,\bm{k}}^{\dagger}, \hat{c}_{+3/2,\bm{k}}^{\dagger},\hat{c}_{-3/2,\bm{k}}^{\dagger} \right)^T$, 
constrained by TRS $\mathcal{T} = \tau_0 (i\sigma_y) \mathcal{K}$ with complex conjugation operator $\mathcal{K}$ and SIS $\mathcal{P} = \tau_z\sigma_0$, the effective Hamiltonian for Dirac semimetal Na$_3$Bi reads
    \begin{equation}\begin{split}
    \mathcal{H}_{\bm{k}}^{D} = & \left( \begin{array}{cccc}
    m_{\bm{k}} & 0 & \hbar v_{\perp} k_+ & 0\\
     & m_{\bm{k}} & 0 & - \hbar v_{\perp} k_-\\
     &  & - m_{\bm{k}} & 0\\
    \dagger &  &  & - m_{\bm{k}}
  \end{array} \right) - \mu,
    \end{split}\end{equation}
where mass term $m_{\bm{k}}=m_{\perp}(k_{x}^{2}+k_{y}^{2})+m_{z}k_{z}^{2}-m_{0}$. Let $\hat{a}^{\dagger}$ and $\hat{a}$ be Landau level creation and annihilation operators with $[\hat{a},\hat{a}^{\dagger}]=1$, the quantization scheme $k_{x}=(\hat{a}+\hat{a}^{\dagger})/\sqrt{2}l_{B}$ and $k_{y}=i(\hat{a}-\hat{a}^{\dagger})/\sqrt{2}l_{B}$
gives the magnetic Hamiltonian,
    \begin{equation}\begin{split}
    \hat{\mathcal{H}}_{k_{z}}^{D} 
    = & \left( \begin{array}{cccc}
    m_{k_{z}} & 0 & \frac{\sqrt{2} \hbar v_{\perp}}{l_B}\hat{a}^{\dagger} & 0\\
     & m_{k_{z}} & 0 & - \frac{\sqrt{2} \hbar v_{\perp}}{l_B}\hat{a}\\
     &  & - m_{k_{z}} &0\\
    \dagger &  &  & - m_{k_{z}}
  \end{array} \right) 
  +\frac{2 m_{\perp}}{l_B^{2}}\left(\hat{a}^{\dagger}\hat{a}+\frac{1}{2}\right)\tau_{z}\sigma_{0} - \mu ,
\end{split}\end{equation}
on the Landau level basis $(|n\rangle,|n-1\rangle,|n-1\rangle,|n\rangle)^T$, where mass term becomes $m_{k_{z}}=m_{z}k_{z}^{2}-m_{0}$. Then effective Hamiltonian for two lowest Landau bands on basis $(|0\rangle,|0\rangle)^{T}$ reads
\begin{equation}\begin{split}
  \mathcal{H}_{0,k_{z}}^{D} 
  = \left(m_{k_{z}}+\frac{m_{\perp}}{l_{B}^{2}}\right)\sigma_z - \mu\sigma_{0},
\end{split}\end{equation}
and all higher Landau bands ($n>0$) are
\begin{equation}\begin{split}
  E^{D}_{\pm n,k_z} 
  = \pm \sqrt{\left[m_{k_{z}}+\frac{2m_{\perp}}{l_{B}^{2}}(n+\frac{1}{2})\right]^{2} 
  + \frac{2 \hbar^2 v_{\perp}^2}{l_B^2} n} - \mu.
\end{split}\end{equation}
When $B=10T$, the ``zero-point'' energy $m_{\perp}/l_{B}^{2}$ is $0.0016eV$, much smaller than $m_{0}=0.086eV$ and negligible in the discussion of zeroth Landau bands.

\begin{figure}[h]\label{FigS1}
  \includegraphics[width=0.6\textwidth]{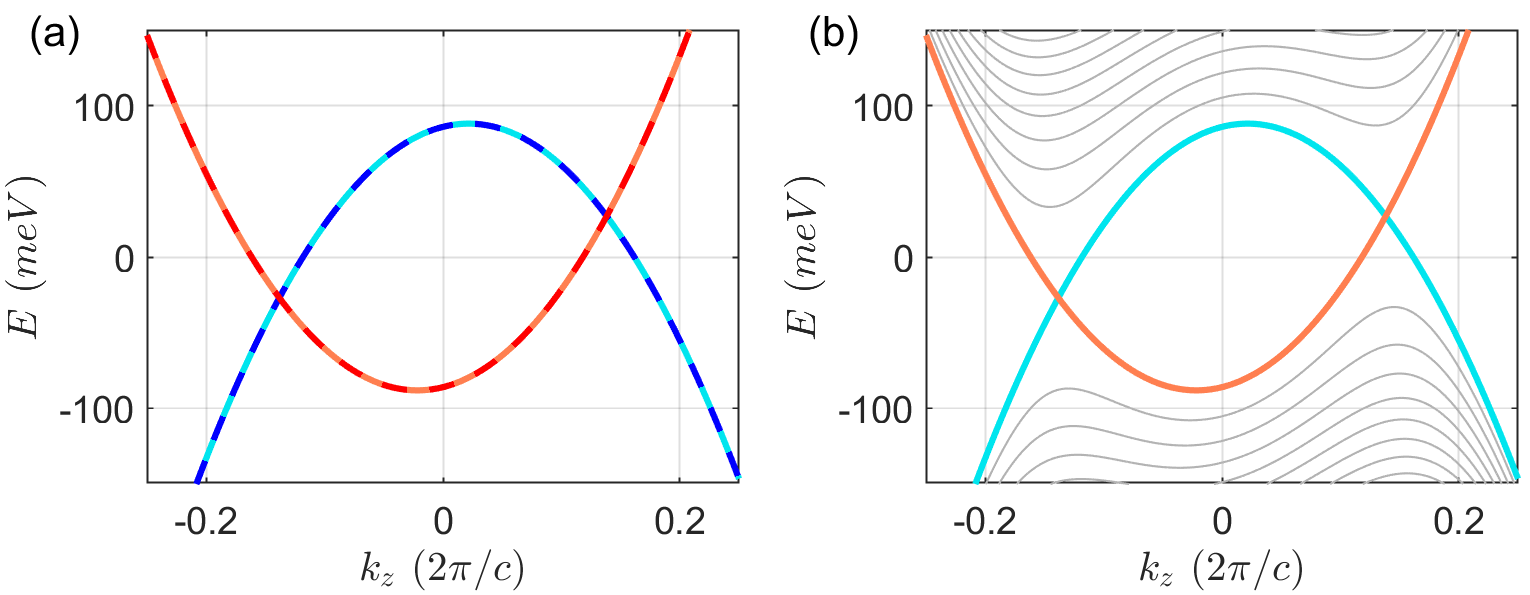}
  \caption{(a) Band structures along $k_{z}$-axis, of Dirac semimetal when space inversion symmetry is broken. (b) Landau bands when a strong magnetic field is applied.}
\end{figure}

As discussed in the maintext, since two vallies are connected by SIS, two degenerate linearly polarized TA phonons will condensate simutaneouly. However, if SIS is broken, two Dirac vallies is no longer degenerate, as shown in Fig.\ref{FigS1}, and the ground state will be STW phase rather than SSW phase when TA phonon beat LA phonon.

\subsection{Kramers-Weyl Semimetal\label{app:C-2}}
    
\begin{figure}[h]\label{FigS2}
    \includegraphics[width=0.98\textwidth]{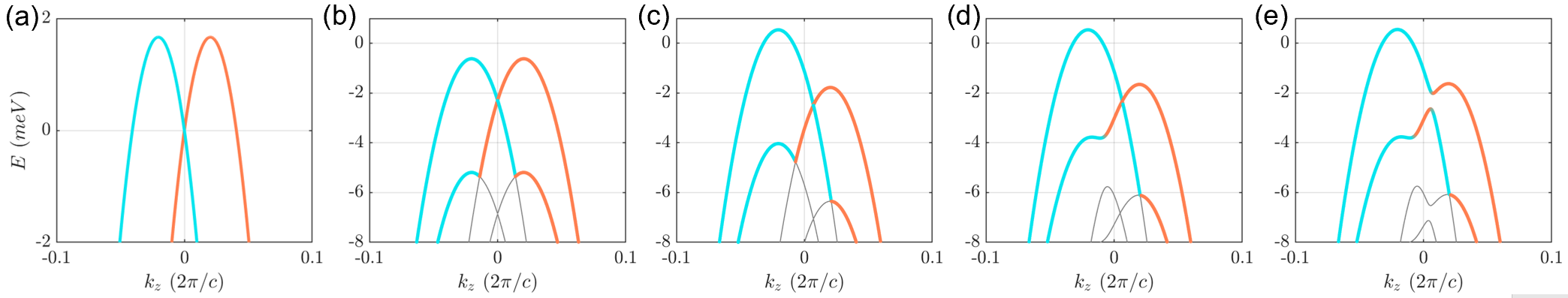}\label{figS2}
    \caption{Evolution of Landau bands of Kramers-Weyl semimetal: (a) before $B_{z}$ applied, (b) only Landau level effect considered, (c) Zeeman effect $\delta_{Z}$ added, (d) in-plane chiral coupling term (SOC) $v_{\perp} \bm{k}_{\bm{\perp}} \cdot \bm{\sigma}_{\perp}$ considered, (d) rotation symmetry breaking term $\delta_{T}$ added.
    }
    \end{figure}

On the basis 
$\hat{\Psi}^{KW\dagger}_{\bm{k}}=\left(\hat{c}_{1/2\bm{k}}^{\dagger},\hat{c}_{-1/2\bm{k}}^{\dagger}\right)^T$, 
the effective Hamiltonian for Kramers-Weyl semimetal is,
\begin{equation}\begin{split}
  \mathcal{H}_{\bm{k}}^{KW} 
  = & (v_{z}k_{z}\sigma_{z} + v_{\perp}\bm{k}_{\bm{\perp}} \cdot
  \bm{\sigma}_{\perp})
  +(u_z k_z^{2} + u_{\perp} \bm{k}_{\perp}^2 - \mu)\sigma_0 .
\end{split}\end{equation}
And its magnetic Hamiltonian is
\begin{equation}\begin{split}
  \hat{\mathcal{H}}_{k_{z}}^{KW} 
  = & \left[v_z k_z\sigma_z + (u_z k_z^2-\mu)\sigma_{0} \right]
     + u_{\perp}\frac{2}{l_B^2}\left(\hat{a}^{\dagger}\hat{a}+\frac{1}{2}\right)\sigma_0
     + v_{\perp}\frac{\sqrt{2}}{l_B}(\hat{a}^{\dagger}\sigma_{-}+\hat{a}\sigma_{+})
     + (-\delta_{Z}\sigma_{z}+\delta_{T}\sigma_{x}),
\end{split}\end{equation}
where $\sigma_{\pm} = (\sigma_x \pm i \sigma_y)/2$, the first and second terms are directly from diagnal elements of Hamiltonian, the third term is SOC, and the fourth term contains Zeeman effects $\delta_{Z}=g_{z}\mu_{B}B_{z}$ in $z$-axis and rotation symmetry breaking term $\delta_{T}$. 
The Fig.\ref{FigS2} is numerically computated on the Landau level basis $(|n\rangle,|n\rangle)^{T}$ with cutoff $n = 10$. To clearly show the influences of each terms, we start with the band dispersion along $z$-axis shown in Fig.5(a), then turn to system under magnetic field with only the first term, and turn on the left terms one by one, to obtain band structures shown in Fig.(c-e).

\subsection{Mean-Field Approximation\label{app:C-3}}
In mean-field level, the phonon creation and annihilation operators are approximated by their expectation values leading to mean-field normal coordinates, then electron-phonon interaction becomes,
\begin{equation}\begin{split}
  \hat{\bar{H}}_{Q}^{ep}
    =&\frac{1}{N}\sum_{mk_{z}}\Delta_{mk_{z}Q}\hat{\Psi}_{mk_{z}Q}^{\dagger}\hat{\Psi}_{mk_{z}Q},\\
  \Delta_{mk_{z}Q}
    =&\sum_{\lambda}\frac{1}{\sqrt{N}}\tau_{+}\mathcal{G}_{mk_{z}Q}^{\lambda}(\langle\hat{b}_{\lambda Q}\rangle+\langle\hat{b}_{\lambda,- Q}^{\dagger}\rangle)+h.c.\\
    = & \frac{1}{\sqrt{N}}\tau_{+}\left(\begin{array}{cc}
      0 & (1-ie^{i\phi_{xy}})\Delta_{Q_{-}}^{-} \\
      (1+ie^{i\phi_{xy}})\Delta_{Q_{+}}^{+}  & 0
    \end{array}\right)
    +\frac{1}{\sqrt{N}}\Delta_{Q_{z}}^{z}\tau_{+}\sigma_{z} + h.c.,\\
    \Delta_{Q_{\pm}}^{\pm}
    =&i2g_{\pm}Q_{\pm}\xi_{\pm,Q_{\pm}}\langle\hat{b}_{\pm,Q_{\pm}}\rangle e^{i\phi_x},\\
    \Delta_{Q_{z}}^{z}
    =&i2g_{z0}Q_{z}\xi_{z,Q_{z}}\langle\hat{b}_{z,Q_{z}}\rangle e^{i\phi_z},\\
    \end{split}\end{equation}
where $\phi_{xy}=\phi_{y}-\phi_{x}$ is the relative phase between TA modes in $x$ and $y$ directions. 
In centrosymmetric case, we have $g_{+}=-g_{-}^{\star}$ and $\hat{\mathcal{P}}\hat{X}_{+,q_{z}}\hat{\mathcal{P}}^{-1}=\hat{X}_{-,q_{z}}^{\dagger}$, Meanwhile, $\Delta_{Q_{T}}^{-}=(\Delta_{Q_{T}}^{+})^{\star}$. By defining $\Delta^{T}_{Q_{T}}=|\Delta_{Q_{T}}^{\pm}|$, up to a unitary transformation, order parameter from electron-TA phonon interaction is simplified to
    \begin{equation}\begin{split}
    \Delta_{mk_{z}Q_{T}}^{T}
    =&\frac{1}{\sqrt{N}}\tau_{+}
     (\sigma_{x}+e^{i\phi_{xy}}\sigma_{y})\Delta_{Q_{T}}^{T}+h.c.,\\
    \end{split}\end{equation}
thus, the phonon energy is
\begin{equation}\begin{split}
  \langle \hat{H}^{ph} \rangle 
  = \frac{1}{N}\sum_{\lambda,q_z}\hbar\omega_{\lambda q_z} 
        \langle\hat{b}^{\dagger}_{\lambda q_z}\hat{b}_{\lambda q_z} \rangle
  = \frac{1}{N}\sum_{\lambda}2\hbar\omega_{\lambda,Q_{\lambda}}
        \langle\hat{b}_{\lambda,Q_{\lambda}}\rangle^{2}
  = M\sum_{\lambda}\left(v_{\lambda}^{ph}|\Delta_{Q_{\lambda}}^{\lambda}| g_{\lambda}^{-1}\right)^{2},
\end{split}\end{equation}
where $\langle \hat{b}_{\mp,Q}\rangle = \langle\hat{b}_{\pm, -Q}^{\dagger} \rangle$ is assumed. 
Now we have the total electronic Hamiltonian in mean-field level,
\begin{equation}\begin{split}
    \bar{\mathcal{H}}_{k_z}^{D} 
    =&(\mathcal{H}_{k_{z}+q_{z}/2}^{D}\oplus \mathcal{H}_{k_{z}-q_{z}/2}^{D}-\mu)
    +
    \tau_{+}[\Delta^{T}(\sigma_{x}+e^{i\phi_{xy}}\sigma_{y})+\Delta^{L}\sigma_{z}]
    +h.c.,\\
    \mathcal{H}_{k_{z}}^{D}
    =&
    \left(\begin{array}{cc}
    E_{-0,k_z}& 0\\
    0 & E_{+0,k_z}
    \end{array} \right),
    \end{split}\end{equation}
for Dirac semimetal and
    \begin{equation}\begin{split}
    \bar{\mathcal{H}}_{k_{z}}^{KW}
    =&(\mathcal{H}_{k_{z}+q_{z}/2}^{KW}\oplus \mathcal{H}_{k_{z}-q_{z}/2}^{KW}-\mu)
    +\tau_{+}[\Delta^{T}(1-e^{i\phi_{xy}})\sigma_{+}+\Delta^{L}\sigma_{z}]+h.c.,\\
    \mathcal{H}_{k_{z}}^{KW}
    =&
    \left(\begin{array}{cc}
    E^{(1)}_{\frac{1}{2},k_z}& \delta_{T}\\
    \delta_{T} &E^{(1)}_{-\frac{1}{2},k_z}
    \end{array} \right),
    \end{split}\end{equation}
for Kramers-Weyl semiemtal with rotation symmetry breaking term $\delta_{T}$.

\section{Renormalized Phonon Dispersion\label{app:E}}

The Mastubara Green's function of ``bare'' phonon is
\begin{equation}\begin{split}
  D_{\lambda\bm{q}}^{(0)} (i \nu_n)
  = \frac{2 \omega_{\lambda\bm{q}}}{(i \nu_n)^2 -(\omega_{\lambda\bm{q}})^{2}} , 
\end{split}\end{equation}
where $\nu_n = 2 \pi n / \beta$ is the Mastubara frequencty. Consider electron-phonon interaction in random phase approximation and self-consistant Midgal approximation, the phonon Green function and phonon self-energy are,
    \begin{equation}\begin{split}
  D_{\lambda\bm{q}}^{-1} 
  = [D_{\lambda\bm{q}}^{(0)}]^{-1}-\Pi^{ph}_{\lambda q_{z}},~~
  \Pi^{ph}_{\lambda q_{z}}=\frac{1}{\hbar}|\mathcal{G}_{\lambda\bm{q}}|^2\mathscr{L}_{\lambda\bm{q}},
    \end{split}\end{equation}
where $\mathscr{L}_{\lambda\bm{q}}(\omega)$ is the Lindhard response function. The pole of $D_{\lambda\bm{q}}(i\nu_{n})$
gives the renormalized phonon frequency $\omega_{\lambda\bm{q}}^{ren}$ in $q_{z}$ direction,
\begin{equation}\begin{split}\label{eq:renormalizedTA}
  \Delta\omega^{2}_{\lambda q_{z}}
  =&(\omega_{\lambda q_{z}}^{ren})^{2} - \omega_{\lambda q_z}^{2}
  =\frac{2}{\hbar}\omega_{\lambda q_z}|\mathcal{G}_{\lambda q_z} |^2\mathscr{L}_{\lambda q_z}(\omega_{\lambda q_{z}}^{ren})
  =\frac{(g_{\lambda}q_{z})^2}{M}\mathscr{L}_{\lambda q_z}(\omega_{\lambda q_{z}}^{ren}), 
\end{split}\end{equation}
where the Lindhard response function are
\begin{equation}\begin{split}
  \mathscr{L}_{\lambda q_z} 
   =&\frac{1}{N_{z}}\sum_{\alpha\beta,k_{z}}\frac{\mathcal{S}_{k_{z}q_{z}}^{\alpha\beta\lambda}(f_{\alpha k_z+q_z} - f_{\beta k_z})}{\varepsilon_{\alpha k_z + q_z} - \varepsilon_{\beta k_z}-\hbar(\omega+i\eta)},\\
  \mathcal{S}_{k_{z}q_{z}}^{\alpha\beta\lambda}
   =&\sum_{ij}\langle s_{z,k_{z}+q_{z}}^{\beta}|s_{z}^{(j)}\rangle\langle s_{z}^{(i)}|s_{z, k_{z}}^{\alpha}\rangle \delta(s_{z}^{(i)}-s_{z}^{(j)}-l_{z,\lambda}^{ph}),\\
\end{split}\end{equation}
where $f_{\alpha k_{z}} = 1 / (e^{\varepsilon_{\alpha k_{z}}/ k_B T} + 1)$ is the Fermi-Dirac distribution function, and $i,j$ run over all possible spin components. When the electronic states are polarized to be independent on crystal momentum, $s_{z,q_{z}}^{\alpha}=s_{z}^{\alpha}$, $\mathcal{S}_{k_{z}q_{z}}^{\alpha\beta\lambda}$ is reduced to the spin selection rule $\delta(s_{z}^{\alpha}-s_{z}^{\beta}-l_{z,\lambda}^{ph})$.

For Na$_{3}$Bi and $\beta$-Ag$_{2}$Se, their bulk modulus (in the sense of Voigt-Reuss-Hill average) $K$ are 17GPa and 58GPa, and their density $\rho$ are 3.63$g/cm^{3}$ and 7.98$g/cm^{3}$. We therefore obtain the Young's modulus $Y=3K(1-2\sigma)$ and LA velocities $v_{s}^{L}=\sqrt{Y/\rho}$ are 2650$m/s$ and 2088$m/s$.

\section{Chiral Standing Wave\label{app:F}}
\begin{figure}[ht]
    \includegraphics[width=0.98\textwidth]{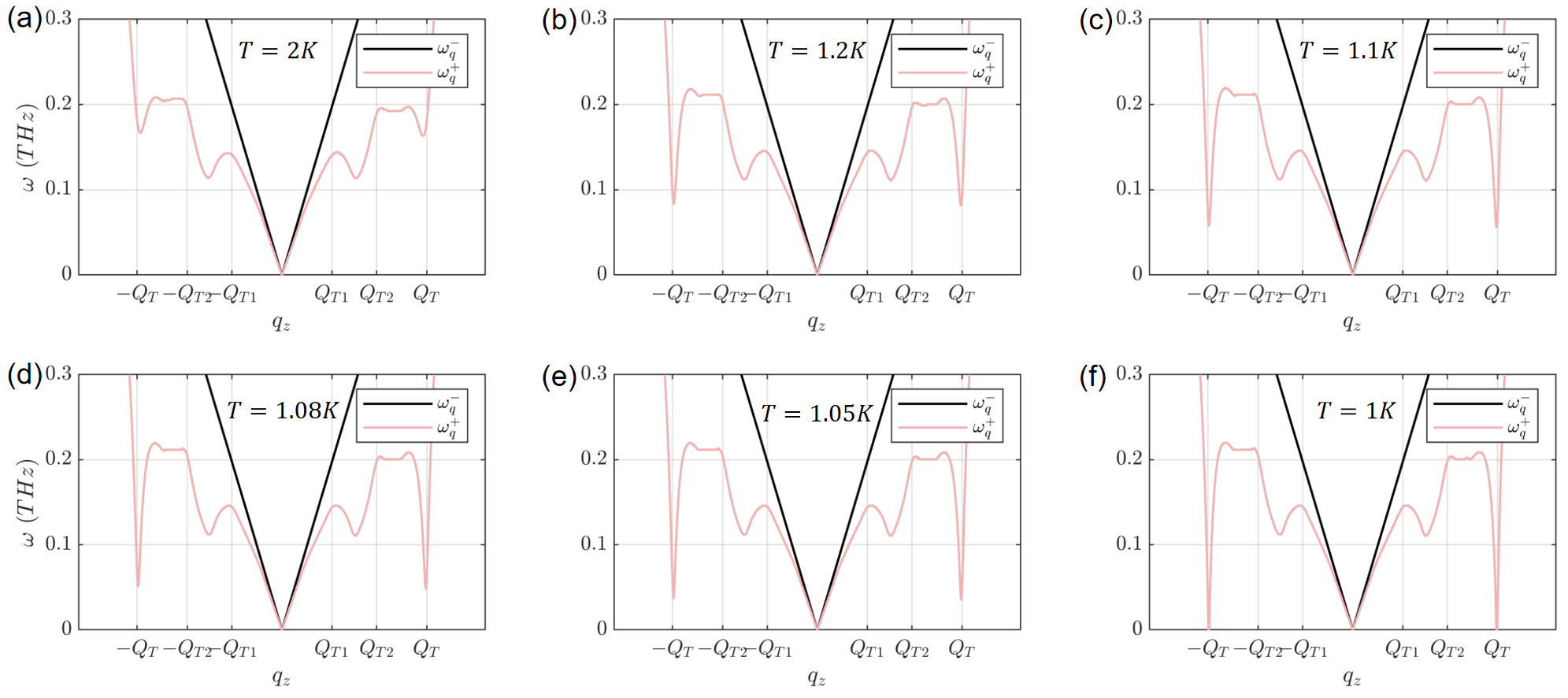}\label{FigS3}
    \caption{Phonon frequencies of Kramers-Weyl semimetal under $B=10T$ magnetic field at different temperatures.
    }
    \end{figure}

At finite temperature, the dressed phonon frequencies from Eq.(\ref{eq:renormalizedTA}) of a generic system have to be solved numerically. To get a flavor of the new standing modes with wave vector around $Q_{T}$, we expand the Lindhard response function up to first order of $\delta q_{z}^{+}=q_{z}-Q_{T}$ and $\omega$,
    \begin{equation}\begin{aligned}
        \mathscr{L}_{\lambda \delta q_{z}^{+}}(\omega)
        =\mathscr{L}_{\lambda}^{(0)}+\mathscr{L}_{\lambda}^{(1)}\delta q_{z}^{+}+\mathscr{L}_{\lambda}^{(2)}\omega.
     \end{aligned}\end{equation}
Then the Eq.(\ref{eq:renormalizedTA}) becomes
    \begin{equation}
        \omega^{2}-\frac{g_{T}^{2}}{M}(\mathscr{L}_{\lambda}^{(0)}+\mathscr{L}_{\lambda}^{(1)}\delta q_{z}^{+}+\mathscr{L}_{\lambda}^{(2)}\omega)(\delta q_{z}^{+}+Q_{T})^{2}
        -(v_{T}^{ph})^2(\delta q_{z}^{+}+Q_{T})^{2}=0,
    \end{equation}
where the non-zero $\mathscr{L}_{\lambda}^{(2)}$ is induced by TRS breaking, and we solve out the effective frequency up to $(\delta q_{z}^{+})^{2}$,
    \begin{equation}\begin{aligned}
    \omega_{\lambda \delta q_{z}^{+}}^{(+Q_{T})}=
     \omega_{Q_{T}}+\frac{1}{2}a(\delta q_{z}^{+})^{2}.\\
    \end{aligned}\end{equation}
As plotted in Fig.(\ref{FigS3}), the parabolic approximation is approximately valid within the $(0.95,1.05)Q_{T}$ regime.  Correspondingly, the discrete wave vectors for standing modes are $\delta q_{n}^{+}=(n+N_{s})\pi/d-Q_{T}\approx n\pi/d(n=0,1,\cdots)$, where $N_{s}$ is a big positive number to cancel the finite $Q_{T}$. Thus, the frequency differences of these standing modes are
    \begin{equation}\begin{aligned}
        &\omega_{n}^{(+Q_{T})} = \omega_{Q_{T}}+\frac{1}{2}a(\delta q_{n}^{+})^{2}.\\
    \end{aligned}\end{equation}
As for the dispersion around $-Q_{T}$, the dispersion could be different from $+Q_{T}$ regime since all time reversal, inversion and rotation symmetries are broken. However, the Fermi surface can be simply treated as two points at low temperature, and the inversion symmetry is approximately recovered. Thus, the effective dispersion near $-Q_{T}$ in terms of $\delta q_{n}^{-}=(n-N)\pi/d-(-Q_{T})\approx n\pi/d$ is
    \begin{equation}\begin{aligned}
    \omega_{n}^{(-Q_{T})} = \omega_{Q_{T}^{-}}+\frac{1}{2}b(\delta q_{-n}^{-})^{2}.
    \end{aligned}\end{equation}
Given a resonant frequency $\omega>\omega_{Q_{T}}$, there are four degenerate left-handed traveling modes, $(\delta q_{n}^{+},\omega_{n}^{(Q_{T})})$, $(\delta q_{-n}^{+},\omega_{-n}^{(Q_{T})})$, $(\delta q_{n}^{-},\omega_{n}^{(-Q_{T})})$, $(\delta q_{-n}^{-},\omega_{-n}^{(-Q_{T})})$, existing to form two left-handed standing modes, and the frequency differences between two adjacent modes are
    \begin{equation}\begin{aligned}
    \Delta\omega_{\pm n}^{(\pm Q_{T})} = \omega_{\pm(n+1)}^{(\pm Q_{T})}-\omega_{-n}^{(\pm Q_{T})}
        =\frac{b\pi^{2}}{d^{2}}\left(n+\frac{1}{2}\right)
        \equiv \Delta\omega_{n}^{(Q_{T})}.
    \end{aligned}\end{equation}
In the case of Ag$_{2}$Se when $T=1.1K$, we have $Q_{T}=6.67\times 10^{8}m^{-1}$, $\omega_{Q}=56.1GHz$, and $a=8.71\times 10^{-4}m^{2}/s$. The $N_{s}$ in this case is $2.12\times 10^{6}$. Given $d=1cm$, we have $\omega_{Q}+\Delta\omega_{n}^{(Q)} =56.1GHz+0.086(n+1/2)kHz$. Meanwhile, we have another long wavelength standing waves, with $\omega_{n'}^{(0)}=nv_{T}^{ph}\pi/d=(268n')kHz$. When $n$ is small, the frequency different between adjacent chiral standing modes indistinguishable compared with long wavelength modes. However, $\Delta \omega_{n}^{(Q_{T})}$ will be even larger than $\Delta\omega_{n'}^{(0)}$ after $n_{c}=3116$ and both of them should be measurable in experiments. As well, $n_{c}/N_{s}\approx 0.15\%$ is a close neighbor of $Q_{T}$ and the parabolic approximation of frequency dispersion still holds. 
    
Because the effective dispersion is sensitive to temperature, we compute the phonon dispersion shown in Fig.(\ref{FigS3}), and the resonant frequencies at different temperatures listed in Table.(\ref{tb:fre}).
\begin{table}[ht]
    \caption{Resonant frequencies of standing modes at different temperatures.}\label{tb:fre}
  \begin{tabular}{c|c|c|c|c}
    \hline\hline
    Temperature ($K$) & $a$ ($m^{2}/s$) & $\omega_{n}^{(0)}$ ($kHz$) & $\omega_{Q_{T}}$ ($GHz$) & $\Delta\omega_{n}^{(Q_{T})}$ ($kHz$) \\
    \hline
    1.2   & $6.17\times 10^{-4}$ & $268n$ & 81.8 & $0.061(n+1/2)$\\
    1.1   & $8.71\times 10^{-4}$ & $268n$ & 56.1 & $0.086(n+1/2)$ \\
    1.08  & $9.80\times 10^{-4}$ & $268n$ & 48.1 & $0.097(n+1/2)$\\
    1.05  & $12.0\times 10^{-4}$ & $268n$ & 35.3 & $0.118(n+1/2)$\\
    \hline\hline
  \end{tabular}
\end{table}

\section{Magneto-acoustic Birefringence\label{app:G}}
\subsection{Faraday Rotation\label{app:G-1}}
Since the right-handed and the left-handed TA acoustic waves propagate with different speeds, $v_{s}^{+}$ and $v_{s}^{-}$, in the gyromagnetic media propagates, the linearly polarized wave becomes
    \begin{equation}\begin{aligned}
         &\bm{X}(d,t)
         =X_{+}e^{i(k_{+}d-\omega t)}\hat{e}_{+}+X_{-}e^{i(k_{-}d-\omega t)}\hat{e}_{-}\\
        =&\frac{X}{2}\left[(e^{ik_{+}d}+e^{ik_{-}d})\hat{\bm{x}}+i(e^{ik_{+}d}-e^{ik_{-}d})\hat{\bm{y}}\right]e^{-i\omega t}\\
        =&Xe^{i[(k_{+}+k_{-})d/2-\omega t]}
         \left[\hat{\bm{x}}\cos\frac{(k_{-}-k_{+})d}{2}+\hat{\bm{y}}\sin\frac{(k_{-}-k_{+})d}{2}\right],
    \end{aligned}\end{equation}
where $|X_{\pm}|=|X_{x}\pm iX_{y}|=X/\sqrt{2}$. The polarization plane therefore rotates by a Faraday angle
    \begin{equation}\begin{aligned}
        \theta_{F}=|\tan^{-1}\frac{X_{y}}{X_{x}}|=\frac{d}{2}|k_{-}-k_{+}|
        =\frac{d}{2}\left|\frac{\omega}{v_{-}^{ph}}-\frac{\omega}{v_{+}^{ph}}\right|
        =\frac{\omega d}{2v_{T}^{ph}}\left(1-\frac{v_{T}^{ph}}{v_{+}^{ph}}\right)
        \equiv \mathcal{V}B_{z}d,
    \end{aligned}\end{equation}
where we used the left-handed TA wave velocity $v_{-}^{ph}=v_{T}^{ph}$, and $\mathcal{V}$ denotes the Verdet coefficient, which is in general a nonlinear function depends on frequency in our case rather than a constant. Then it is estimated that, we numerically obtain $v_{+}^{ph}=1811m/s$ (another possibility $-1811m/s$ is dropped) for $\omega=\omega_{Q}=58.0GHz$ when $T=1.1K$, $|g_{T}/g_{z0}|=1.43\times 10^{-4}$, $B_{z}=10T$ and $\delta_{T}=1.2meV$ for Kramers-Weyl semimetal $\beta$-Ag$_{2}$Se system. 
Consider that $v_{T}^{ph}=v_{z}^{ph}\sqrt{(1+\sigma)/(2-2\sigma)}=852m/s$, we find a huge $\theta_{F}/d=1.77\times 10^{5}rad/cm$.

\subsection{Kerr Ellipticity in Polar Configuration\label{app:G-2}}
In optics, according to the relative configuration of incident light and magnetic materials, there are three types magneto-optical Kerr effect: polar, longitudinal and transverse. Similarly, we also have three configurations in acoustics. For simplicity, here we only take the polar configuration at near normal incidence as an example since it has larger effect than other two. The reflection coefficients for two chiral acoustic waves are obtained through Fresnel relations:
    \begin{equation}\begin{aligned}
        r_{+}=\left|\frac{1-n_{+}/n_{T}}{1+n_{+}/n_{T}}\right|=0.262,~~
        r_{-}=\left|\frac{1-n_{-}/n^{T}}{1+n_{-}/n_{T}}\right|=0.109,
    \end{aligned}\end{equation}
where we used $n^{-}/n^{T}=v_{T}^{ph,CPVC}/v_{T}^{ph,KW}=1060/852=1.244$ and $n^{+}/n^{T}=v_{s}^{ph,CPVC}/v_{+}^{ph,KW}=1060/1811=0.585$. 
The left-handed TA wave will completely incident into the gyromagnetic material, while a proportion of righ-hand rotating wave can reflect with non-zero reflection coefficient $r_{+}$. So the Kerr ellipticity $\epsilon^{K}=(r_{+}-r_{-})/(r_{+}+r_{-})=0.412$.

\end{document}